\documentclass[11pt]{article}
\pdfoutput=1
\usepackage{jheppubme}
\usepackage[table]{xcolor}
\usepackage{graphicx, psfrag, dsfont, amsfonts}
\usepackage{float}
\usepackage{caption, comment}
\usepackage[mathscr]{euscript}
\usepackage{dynkin-diagrams}
\usepackage{appendix}
\usepackage{coolstr}
\usepackage{hyperref}
\usepackage[normalem]{ulem}
\hypersetup{pdfauthor={Name}}
\usepackage{color}
\usepackage{chngcntr}
\counterwithin{table}{subsection}
\usepackage{booktabs,caption}
\usepackage{lscape}
\definecolor{Mygrey}{gray}{0.8}
\definecolor{Mywhite}{gray}{1.0}

\usepackage{amsmath,amssymb,epsfig,jheppubme}
\usepackage{multirow,longtable,enumerate,bm}
\usepackage[flushleft]{threeparttable}

\usepackage{tikz}
\usetikzlibrary{decorations.pathmorphing,arrows.meta}

\font\cmss=cmss12

\global\interfootnotelinepenalty=1000

\newcommand\half{\frac12}

\newcommand\bi{\begin{itemize}}
\newcommand\ei{\end{itemize}}

\newcommand\tk{{\tilde k}}

\newcommand\tf{{\tilde f}}

\newcommand\bea{\begin{eqnarray}}
\newcommand\eea{\end{eqnarray}}
\newcommand\be{\begin{equation}}
\newcommand\ee{\end{equation}}

\newcommand{\cO}{{\cal O}}

\newcommand\mA{{\mathsf A}}

\newcommand\mD{{\mathsf D}}
\newcommand\mE{{\mathsf E}}
\newcommand\mF{{\mathsf F}}
\newcommand\mG{{\mathsf G}}

\newcommand\talpha{{\tilde\alpha}}

\newcommand\sfrac[2]{{\textstyle\frac{#1}{#2}}}

\newcommand\ZZ{\hbox{Z\kern-.4emZ}}
\newcommand\sZZ{\hbox{\sevenfont Z\kern-.4emZ}}

\newcommand{\eref}[1]{Eq.\,(\ref{#1})}
\newcommand{\Comment}[1]{{}}

\usepackage{xparse}
 \NewDocumentEnvironment{spl}{b}{%
   \begin{equation}
     \begin{split}
       #1
     \end{split}
   \end{equation}
 }
 
\def\IB{\relax{\rm I\kern-.18em B}}
\def\IC{{\relax\hbox{\kern.3em{\cmss I}$\kern-.4em{\rm C}$}}}
\def\ID{\relax{\rm I\kern-.18em D}}
\def\IE{\relax{\rm I\kern-.18em E}}
\def\IF{\relax{\rm I\kern-.18em F}}
\def\II{\relax{\rm I\kern-.18em I}}

\def\Id{\relax{1\kern-.32em 1}}
\def\IG{\relax\hbox{$\inbar\kern-.3em{\rm G}$}}
\def\IR{\relax{\rm I\kern-.18em R}}

\title{Unlocking the Wronskian Tower: A Simplification of the Holomorphic Modular Bootstrap} 
\author[a,b]{Arpit Das}
\author[c]{and Sunil Mukhi\,\footnote{Adjunct Professor, ICTS Bengaluru.}} 

\affiliation[a]{School of Mathematics and Maxwell Institute for Mathematical Sciences,\\ University of Edinburgh, Edinburgh EH9 3FD, U.K.}
\affiliation[b]{Higgs Centre for Theoretical Physics,
University of Edinburgh,\\ Edinburgh EH8 9YL, U.K.}
\affiliation[c]{Indian Institute of Science Education and Research,\\ Homi Bhabha Rd, Pashan, Pune 411 008, India}

\emailAdd{arpit.das@ed.ac.uk}
\emailAdd{sunil.mukhi@gmail.com}

\abstract{
Characters of rational conformal field theories solve modular linear differential equations labelled by their order and the Wronskian index $\ell$. Direct classification of admissible solutions by solving MLDEs becomes increasingly difficult at higher $\ell$ -- where movable poles and accessory parameters appear. In this work we introduce differential operators that relate higher-$\ell$ solutions to lower-$\ell$ ones while preserving modular covariance and integrality of the \(q\)-series. In rank two, this generates all allowed Wronskian sectors from the Mathur--Mukhi--Sen equation. In rank three and higher, it reduces the construction of higher-$\ell$ quasi-characters to simpler equations with lower $\ell$. This gives an efficient new route for organising candidate RCFT characters, and more generally quasi-characters, across the Wronskian tower. As an application, we apply our construction to prove a previously conjectured property on the signs of $\ell=2$ quasi-characters in rank 2.
}

\preprint{}

\begin{document}

\maketitle


\section{Introduction}

Rational conformal field theories provide one of the most highly constrained classes of non-trivial field theories, and exact solutions for their partition functions and correlation functions are known in many cases. Despite this, they remain challenging. Their complete classification remains not just unknown, but -- for good reasons -- is considered impossible \cite{DiFrancesco:1997nk}. The most explicit results to date describe the classification of theories with a small number of primaries and a bounded central charge \cite{Mukhi:2022bte, Rayhaun:2023pgc} (these references also discuss why the general problem is essentially intractable).  

A basic reason for their solvability is that their characters form finite-dimensional vector-valued modular forms (VVMF) under the action of \(\rm SL(2,\mathbb Z)\). Thus, if an RCFT has \(p\) independent characters, these characters may be assembled into a \(p\)-component vector
\be
        \chi(\tau)=\big(\chi_0(\tau),\chi_1(\tau),\ldots,\chi_{p-1}(\tau)\big) ,
\ee
whose modular transformation properties encode the modular representation of the theory. 

This observation lies at the heart of the holomorphic modular bootstrap: instead of constructing the full chiral algebra directly, one classifies possible character vectors by requiring modular covariance and physically admissible \(q\)-expansions. The former property is ensured by making them solve a Modular Linear Differential Equation (MLDE) \footnote{Note that solving an MLDE does not ensure absence of logarithmic terms in $q$, nor does it ensure that the \(\rm SL(2,\mathbb Z)\) representation is irreducible. In the present work we consider only non-logarithmic solutions with irreducible modular representation.} \cite{Anderson:1987ge, Mathur:1988na}. Each MLDE has an associated integer, the Wronskian index, -- which captures the pole structure of the coefficient functions. This approach to classification has been rather successful for low numbers $p$ of independent characters (equal to the rank of the MLDE\,\footnote{Throughout the paper, we use {\it rank} to label the order of the MLDE (equivalently, the rank of the VVMF made up of independent components) \cite{Gannon:2013jua}. This is not always equal to the rank of the corresponding Modular Tensor Category \cite{rowell2009classification} for an RCFT, which is the number of primaries rather than independent characters. The former is often lesser than the latter due to the presence of internal symmetries -- for example, automorphisms of the associated  Dynkin diagrams in Wess-Zumino-Witten theories, which cause more than one primary to have the same character.}) and low values of the Wronskian index $\ell$ \cite{Mathur:1988na, Mathur:1988gt, Mathur:1988rx, Harvey:2018rdc, Kawasetsu:2018tzs, Arike:2018fru, Chandra:2018pjq, Bae:2018qym, Mukhi:2020gnj, Grady_2020, franc2020classification, Kaidi:2020ecu, Bae:2020xzl, Kaidi:2021ent, Das:2021uvd, Bae:2021mej, Franc:2021hwo, Kaidi:2022sng, Mukhi:2022bte, Duan:2022kxr, Lee:2023owa, Pan:2023jjw, Das:2022uoe, Rayhaun:2023pgc, Govindarajan:2025rgh, Govindarajan:2025jlq, Govindarajan:2026frs}. The pair \((p,\ell)\) therefore provides a natural label for MLDEs and their solutions.

For any fixed $p$, the MLDE is different for each value of $\ell$. For $p=2$, some concrete results for $\ell=6,12$ were found in \cite{Das:2023qns}, but beyond that the approach of directly solving MLDE has not yielded any results with $\ell\ge 6$\,\footnote{The quasi-character approach proposed in \cite{Chandra:2018pjq} has worked better, though it is only complete at $p=2$ with partial results at $p=3$ \cite{Mukhi:2020gnj}.}. For \(\ell<6\), the MLDEs are essentially rigid: their parameters are fixed by the exponents of the corresponding characters, or equivalently by the central charge and conformal dimensions. Starting at \(\ell=6\), new non-rigid data appear. These are associated with ``movable zeroes'' of the Wronskian and accessory parameters of the MLDE. Furthermore, demanding regularity at the elliptic points $\tau= i, \omega$ imposes additional constraints between the movable poles and the accessory parameters. This makes a direct classification by solving MLDEs increasingly difficult.

In this note we take a major step towards overcoming this problem by introducing a simple differential operator that relates MLDEs of different $\ell$-values to each other. This operator, which will play a central role in what follows, is:
\be\label{eq:sc_op}
        \Theta \equiv \eta^{-4}D ,
\ee
where \(\eta\) is the Dedekind eta-function and \(D\) is the Ramanujan--Serre derivative that will be defined precisely below. It is known that $D$ raises modular weight by two and \(\eta^{-4}\) has modular weight \(-2\), so the operator \(\Theta\) preserves the weight of the objects on which it acts, in particular mapping weight-zero objects to weight-zero objects. However it changes the multiplier system $\rho$ of the associated modular transformations in a simple and controlled way\,\footnote{This operator is similar in spirit to those introduced in \cite{Bantay:2010uy, nagatomo2022MLDE}, but the ones used in these references do not introduce a multiplier system and are therefore not as useful for the present purpose as the one we discuss.}. 

This simple differential map greatly simplifies the study of MLDEs. For a fixed number $p$ of characters, rather than treating the MLDE at each value of \(\ell\) as an independent object, one uses the operator \(\Theta\) as a bridge between different values of the Wronskian index. As we will see, this operator acting on rank-$p$ solutions sends $\ell$ to $\ell+p$. Thus high-\(\ell\) solutions can be generated from lower-\(\ell\) data by applying a universal differential operator in suitable combinations with other modular functions. This construction also acts naturally on the larger space of quasi-characters, where the coefficients in the \(q\)-series are rational with bounded denominators, but need not be non-negative.

We first develop this idea in detail for rank two. Starting from solutions of the \((2,0)\) Mathur--Mukhi--Sen equation \cite{Mathur:1988na, Mathur:1988gt}, a single application of \(\Theta\) maps them to solutions of the \((2,2)\) MLDE studied in \cite{Naculich:1988xv, Hampapura:2015cea, Gaberdiel:2016zke}. This map is also invertible showing that the relation gives a one-to-one correspondence between the solution spaces. For two characters, repeated applications of \(\Theta\) raise the Wronskian index by two at each step. This is very convenient since it was already shown in \cite{Naculich:1988xv} that the Wronskian index for two characters is even. Thus we actually obtain {\em all} allowed Wronskian indices in this case.

After discussing the rigid cases of \(\ell=2\) and \(\ell=4\), we move on to the first non-rigid case at \(\ell=6\). In this case, one finds that the modular functions: \(j^{1/3}\) and \((j-1728)^{1/2}\) can be combined with $\Theta$ in a suitable way determined by the matching of multiplier systems. We show that this introduces precisely one new free parameter, agreeing with the fact that solutions of the $(2,6)$ have one free parameter \cite{Chandra:2018pjq} that is associated to the position of a movable pole \cite{Das:2023qns}. We then extend this result to the case of $(p,\ell)=(2,6r)$ for all positive integers $r$. 

An important feature of the $\Theta$ action that emerges here is that integrality (in the bounded-denominator sense) is preserved: the modular functions used have integral \(q\)-series, while differentiation via the $\Theta$ operator brings down rational factors from the leading exponents -- which again have bounded denominators. Note, however, that the map does not preserve positivity -- hence admissible characters can be mapped to quasi-characters and vice-versa. In fact, this is generically the case.

We then discuss the rank three case. Here each action of $\Theta$ increases $\ell$ by 3. Now it is known \cite{Kaidi:2021ent} that the Wronskian index of 3rd order MLDEs is a multiple of three, hence repeated application of \(\Theta\) (together with the modular functions mentioned above) again generates {\em all} higher-\(\ell\) three-character families from the \((3,0)\) data. A complete classification of rank-three quasi-characters would therefore reduce, through the present construction, to a classification of the \((3,0)\) quasi-characters -- which is already a remarkable simplification. In fact the admissible characters at $(3,0)$ have been classified in \cite{Kaidi:2021ent, Das:2021uvd, Bae:2021mej} and the subset describing CFTs were found in \cite{Das:2022uoe}, while a partial classification of quasi-characters has been presented in \cite{Mukhi:2020gnj} and completing it seems quite a tractable problem.

We then move on to the cases of $p\ge 4$ where the map is still useful but not complete: the $\Theta$ operator does not generate all allowed Wronskian indices starting from $\ell=0$. Even so, the construction remains valuable: it gives explicit families of higher-index MLDE solutions, preserves modular covariance and bounded-denominator integrality, and supplies a systematic alternative to solving complicated higher-\(\ell\) MLDEs.

In the concluding sections we comment on the relation between our construction and Bantay-Gannon duality which also happens to relate $(3,0)$ to $(3,3)$ quasi-characters and was recently discussed in \cite{Govindarajan:2025jlq}. We also compare the $\Theta$ operator to other operators studied in the literature on VVMFs previously in \cite{Bantay:2010uy, nagatomo2022MLDE}.

\section{Background and review} \label{sec:bckgrd_11}

\subsection[MLDE and valence formula]{MLDE and valence formula\,\footnote{The reader familiar with MLDEs may skip this sub-section.}}

\label{sec:bckgrd}

We first introduce the Ramanujan-Serre derivative acting on weight-$k$ modular functions:
\be
D_k \equiv q\frac{d}{dq}-\frac{k}{12}E_2(q)
\ee
After $D_k$ acts, the result is a modular function of weight $k+2$. Hence, as is standard in the literature, the operator $D^p$ acting on weight-$k$ modular functions is taken to mean:
\be
D^p \equiv D_{k+2p-2}\circ D_{k+2p-4}\circ\ldots\circ D_{k+2}\circ D_k
\label{Dpowerp}
\ee
The above notation allows us to drop the subscript $k$ on $D$ as long as the weight of the object on which it acts is clear from the context. In particular RCFT characters have weight 0, so in that context the $p$-th derivative is the operator in \eref{Dpowerp} with $k=0$.

The generic $p$-th order MLDE with arbitrary Wronskian index $\ell$ may be written:
\be
\Big(D^p+\sum_{s=1}^p \mu_{2s}\,\phi_{2s} D^{p-s}\Big)\chi=0
\label{geneqn}
\ee
Here the $\mu_{2s}$ are arbitrary real parameters, while the $\phi_{2s}$ are meromorphic modular forms of weight $2s$. Their pole structure is determined by the zeroes of the Wronskian, and their normalisations are chosen such that the leading coefficient is unity. Explicitly, we have:
\be
\mu_{2s}\,\phi_{2s}=(-1)^{s}\frac{W_{p-s}}{W_p}
\label{phidef}
\ee
where the generalised Wronskians are defined as:
\be
W_{s}(\tau)\equiv \left| 
\begin{matrix}
\chi_0 & \chi_1& \cdots & \chi_{p-1}\\
\vdots & \vdots & \vdots & \vdots \\
D^{s-1}\chi_0 & D^{s-1} \chi_1 & \cdots & D^{s-1}\chi_{p-1}\\
D^{s+1}\chi_0 & D^{s+1} \chi_1 & \cdots & D^{s+1}\chi_{p-1}\\
\vdots & \vdots & \vdots & \vdots \\
D^{p}\chi_0 & D^{p} \chi_1
& \cdots & D^{p}\chi_{p-1}
\end{matrix}
\right|
\label{Wronskians}
\ee

Now it is easy to see from the definition that $W_{p-1}=D W_p$ and using \eref{phidef} we find the nice relation:
\be
\mu_2\phi_2=-\frac{W_{p-1}}{W_p}=-D\log W_p
\label{phitwoprop}
\ee

The $p$ independent solutions of \eref{geneqn} admit $q$-expansions as follows
\be
\chi_i(q)=q^{\alpha_i}\sum_{n=0}^{\infty}a_{i,n}q^n,
\qquad i=0,1,\ldots,p-1.
\label{qseries}
\ee
The solutions are holomorphic in the upper half-plane and may have a pole at the cusp $\tau\to i\infty$.

Now if we write: $\alpha_{i}=-\frac{c}{24}+h_i$ with $h_0 = 0$, then the valence formula for the Wronskian is \cite{Mathur:1988na}:
\be
\sum_{i=0}^{p-1} \alpha_i = -\frac{pc}{24}+\sum_{i=0}^{p-1}h_i = \frac{p(p-1)}{12}-\frac{\ell}{6}.
\label{valence}
\ee
and this determines $\ell$ in terms of $p,c,h_i$. When the solutions are CFT characters, $c$ and $h$ are the chiral central
charge and the conformal dimension; otherwise, they are simply labels for the leading exponents of the solutions.

In the present work we are interested in all solutions of \eref{geneqn} that have an integrality property: the coefficients $a_{i,n}$ in the $q$-series \eref{qseries} are all rational with bounded denominators. Thus, after an overall normalisation by an integer (the largest denominator) the coefficients all become integral. They need not, however, be positive. Among these, the {\it admissible characters} are those where the coefficients are all non-negative and one of them has integral coefficients upon normalising the leading term to unity\,\footnote{The latter requirement reflects a non-degenerate vacuum for the theory.}. The remaining ones are known as {\it quasi-characters} \cite{Chandra:2018pjq}.

\subsection{Quasi-characters in rank 2} \label{sec:quasi2}
Let us first review quasi-characters for rank-2 MLDEs. Complete results for quasi-characters are known only in the case of two-component VVMFs which solve second-order MLDEs. The most general such MLDE may be written as \cite{Mathur:1988na}\,\footnote{Wherever needed, we denote the characters as $\chi^{(\ell)}_i$ and the corresponding central charge and conformal dimensions as $c^{(\ell)}$ and $h^{(\ell)}_i$ where $\ell$ in the superscript denotes the associated Wronskian index.}:
\be
\Big(D^2+\sfrac{\ell}{6}\phi_2D
+\phi_4\Big)\chi^{(\ell)}=0 ,
\label{gentwo}
\ee
whose solutions obey the valence formula obtained by specialising \eref{valence} to $p=2$:
\be
\ell=1-6h^{(\ell)}+\frac{c^{(\ell)}}{2}.
\label{valence.ch}
\ee
For second-order MLDEs, the Wronskian index is necessarily even
\cite{Naculich:1988xv}.

To find quasi-character solutions we start with the simplest case of vanishing Wronskian index, $\ell=0$ in \eref{gentwo}. There is no weight-2 holomorphic modular form $\phi_2$, and the Eisenstein $E_4$ is the unique one of weight 4 up to a scalar multiple that we denote $\mu$. Thus we get the MMS equation \cite{Mathur:1988na}:
\be
\left(D^2+\mu E_4\right)\chi^{(0)} =0,
\qquad
\mu=-\frac{c^{(0)}(c^{(0)}+4)}{576}\\[2mm]
\label{2ell0}
\ee
This equation has infinitely many quasi-character solutions that fall into seven distinct families\footnote{Here for simplicity of exposition we are consistently ignoring three more families, two of which are associated to non-unitary CFT at best, and one collapses to single-character solutions that are modular invariant -- up to a phase -- on their own.}. For each solution, one member
of the pair has all positive coefficients, while the other has coefficients of alternating sign up to some term in the series after which they acquire a definite asymptotic sign. Choosing each solution to start with a positive term, we call the alternating solution {\it Type~I} when the asymptotic sign is positive and {\it Type~II} when it is negative \cite{Chandra:2018pjq, Das:2025gto}. 

The seven families can be parametrised by their central charge \cite{Das:2025gto}
\be
c^{(0)}=24M^{(0)}+j^{(0)},
\label{cqcell.0}
\ee
where the family is labelled by:
\be
j^{(0)}=1,2,\frac{14}{5},4,\frac{26}{5},6,7.
\label{jvalell.0}
\ee
while $M^{(0)}$ labels the different members of each family.

Using $\ell=0$ in \eref{valence.ch}, one finds
\[
h^{(0)}=2M^{(0)}+\frac{j^{(0)}+2}{12},
\]
and hence
\be
\begin{split}
\alpha_0^{(0)}=-M^{(0)}-\frac{j^{(0)}}{24},\quad \alpha_1^{(0)}=M^{(0)}+\frac{j^{(0)}+4}{24}.
\end{split}
\label{exps.0}
\ee
The transformation $c\leftrightarrow-c-4$ exchanges $\alpha_0$ and
$\alpha_1$, and therefore interchanges the two associated solutions.

At $M^{(0)}=0$, one recovers the seven admissible MMS solutions. These correspond to the WZW models $\mA_1, \mA_2, \mG_2, \mD_4, \mF_4, \mE_6, \mE_7$. All other members of each family have some negative coefficients and correspond to quasi-characters. We will use the Kac-Moody algebra names above to label the corresponding infinite families. 

Next consider the $\ell=2$ specialisation of \eref{gentwo}.
\be
\left(D^2+\frac13 \frac{E_6}{E_4}D +\mu E_4\right)\chi^{(2)} = 0,\qquad \mu=-\frac{c^{(2)}(c^{(2)}-4)}{576},
\label{2ell2}
\ee
This also has seven families of quasi-character solutions\,\footnote{ignoring three more classes with special behaviour, as we did in the $\ell=0$ case} as was first observed in \cite{Hampapura:2015cea}. These were identified as cosets of a meromorphic CFT with $c=24$ by the unitary level-one WZW models obtained as MMS solutions \cite{Gaberdiel:2016zke}.

This time the valence formula reads:
\be
c^{(2)}-2=12h^{(2)}
\label{valence.2}
\ee
and the exponents of the two solutions are parametrised as:
\be
\begin{split}
\alpha_0^{(2)} = -\frac{c^{(2)}}{24},\qquad
\alpha_1^{(2)} = -\frac{c^{(2)}}{24}+h^{(2)}=\frac{c^{(2)}-4}{24}
\end{split}
\label{alphaelltwo}
\ee
The central charges now are:
\be
c^{(2)}=24M^{(2)}+j^{(2)}
\label{tcqcell.2}
\ee
where:
\be
j^{(2)}=23, 22, \frac{106}{5}, 20, \frac{94}{5}, 18, 17
\label{tjvalell.2}
\ee
The $M^{(2)} = 0$ member of each sub-series, with $c^{(2)}=j^{(2)}$, is admissible and corresponds to one or more RCFTs that can be understood as cosets of a meromorphic CFT \cite{Schellekens:1992db} by one of the MMS theories. This was first explained in \cite{Gaberdiel:2016zke}.

A central result of \cite{Chandra:2018pjq} is that quasi-characters with Wronskian indices $\ell=0,2,4$ generate admissible VVMFs with indices $6r,6r+2,6r+4$, respectively, for every non-negative integer $r$ and moreover {\em every} such VVMF can be obtained in this way. Examples
realised by actual CFTs were identified in \cite{Chandra:2018ezv, Das:2023qns}. Since the MLDEs with $\ell<6$ can be analysed explicitly, this reduces the construction of arbitrary-index admissible characters to a small set of low Wronskian index quasi-characters.

Now, in \cite{Gaberdiel:2016zke} it was observed that the $\ell=0$ and $\ell=2$ admissible solutions obey a bilinear  coset-type relation. But this in fact extends more generally to pairs of quasi-characters, with each $\ell=0$ solution having a unique $\ell=2$ dual partner.  In the present work we will find a different pairing between $\ell=0,2$ quasi-characters that uses our $\Theta$-map  \eref{eq:sc_op}. In more general cases, the latter map will work even in cases where bilinear pairings are not known.

\subsection{Quasi-characters in rank $\ge 3$}

The case of rank-3 quasi-characters was first investigated in \cite{Mukhi:2020gnj} for the $\ell=0$ case. The idea was to take known rank-3 admissible solutions that were known at the time \cite{Mathur:1988gt, Gaberdiel:2016zke, Hampapura:2016mmz,  franc2016hypergeometric} and try to guess infinite families, each one including one of the admissible solutions, and verifying the desired integrality properties. This in turn relied on classifications of fusion classes with three or more primary fields \cite{Christe:1988xy}. Over a dozen infinite classes of quasi-character solutions were found and are listed in Tables 1,2 of \cite{Mukhi:2020gnj}. 

Subsequently, additional admissible rank-3 solutions at $\ell=0$ were found in \cite{Kaidi:2021ent, Das:2021uvd, Bae:2021mej}. Moreover the $\ell=3$ case was investigated in \cite{Gowdigere:2023xnm}. Finally, in \cite{Govindarajan:2025rgh, Govindarajan:2025jlq}, the techniques of Bantay-Gannon theory \cite{Bantay:2005vk, Bantay:2007zz} were combined with MLDE methods. This powerful approach led to both admissible characters and quasi-characters being found for $\ell=0,3$ and admissible solutions for higher $\ell$--values as well. Despite all this, there is no complete classification of quasi-characters in rank-3, even at $\ell=0$, in the form of families analogous to Eqs. (\ref{cqcell.0}) and (\ref{jvalell.0}) such that linear superpositions within each family generate all admissible characters.

The rank-3 case presents significant differences from rank-2. For one, there are infinitely many {\em admissible} characters in this case \cite{Mathur:1988gt, Das:2021uvd}, and one expects each one in turn to lie in an infinite family of quasi-characters. For another, there are three independent ways to choose an ``identity character'', generalising the exchange property mentioned below \eref{exps.0} in rank 2.

For higher rank, a few examples at rank 4, 5 and 6 were constructed in \cite{Govindarajan:2025rgh}. We are not aware of any results on quasi-character solutions of MLDE beyond the above.

\section{Relating rank-2 MLDEs by the $\Theta$-map}

In this section we describe the action of the $\Theta$-map on generic rank $2$ MLDEs. We begin with the simplest examples and then move on to the most general ones.

\subsection{Wronskian index $0\le \ell < 6$}\label{sec:rk2}

We start with the MMS equation \eref{2ell0}. Suppose $\chi^{(0)}$ is a solution of \eref{2ell0}. Then, we show that:
\be\label{eq:rel_20-22}
\chi^{(2)} \equiv \Theta\chi^{(0)}
\ee
is a solution of \eref{2ell2} where for convenience we recall that:
\be
\Theta \equiv \eta^{-4} D
\ee

We have:
\begin{equation}
\begin{split}
D^2\chi^{(2)} & = \eta^{-4}D^3\chi^{(0)}\\
&= \eta^{-4}D(-\mu E_4\chi^{(0)})\\
&= -\mu\eta^{-4}\left(-\frac13 E_6\chi^{(0)}+E_4D\chi^{(0)}\right)\\
&=-\frac{1}{3}\frac{E_6}{E_4}D\chi^{(2)}-\mu E_4\chi^{(2)}
\end{split}
\end{equation}
where in the second line we used the MMS equation \eref{2ell0} and in the third line, carried out the differentiation by $D$. The result shows that $\chi^{(2)}$ indeed satisfies \eref{2ell2}. 

This map can be inverted (upto a non-zero normalisation). We claim that if $\chi^{(2)}$ is a solution of \eref{2ell2} then:
\be
\chi^{(0)} =\eta^4 E_4^{-1}D\chi^{(2)}
\ee
is a solution of \eref{2ell0}. The proof is as follows:
\be
\begin{split}
D^2\chi^{(0)} &=\eta^4D^2(E_4^{-1}D\chi^{(2)})\\
&= \eta^4 D\left(\frac13 \frac{E_6}{E_4^2}D\chi^{(2)}+E_4^{-1}D^2\chi^{(2)}\right)\\
&= -\mu\eta^4 D\chi^{(2)}\\
&= -\mu E_4\chi^{(0)},
\end{split}
\ee
which proves the claim.

We could have predicted without this calculation that $\chi^{(2)}$ obtained as above would be an $\ell=2$ solution. For this, note first that $\Theta$ has the modular transformation properties:
\be
\begin{split}
\Theta(\tau+1)&=e^{-\frac{2\pi i}{6}}\Theta(\tau), \qquad \Theta\left(-\frac{1}{\tau}\right)=-\Theta(\tau)
\end{split}
\label{Thetatransf}
\ee
where the second equation comes from the fact that $\Theta$ has a modular weight 0 (due to cancellation of the weights $+2$ from $D$ and $-2$ from $\eta^{-4}$) but inherits a multiplier system from $\eta(\tau)$:
\be
    \eta(\gamma\tau) = v_\eta(\gamma)(c\tau + d)^{\frac12}\eta(\tau), \qquad \gamma\in\text{SL}(2,\mathbb{Z}),
\label{eta-mod}
\ee
Thus the inherited multiplier system Eq.~\eqref{Thetatransf} is $v_\eta^{-4}$. Let $\rho(\gamma)$ be the modular representation for the $\chi^{(0)}$ VVMF. Then the VVMF obtained after acting with the $\Theta$-map has the modular representation:
\be
    \rho_\Theta(\gamma) = v_\eta^{-4}(\gamma)\otimes\rho(\gamma),
\ee

From the above discussion, we see that if $\chi^{(0)}$ is a weight-0 VVMF then so is $\Theta\chi^{(0)}$, just with different $S$ and $T$ matrices, where the $T$-matrix has additional phases and the $S$ matrix flips its sign. Now let us look at the $q$-expansion. Acting on a weight-0 VVMF, $D$ is just $q\frac{d}{dq}$ and hence preserves the leading power of $q$ in $\chi^{(0)}$ (unless the leading term is constant, which does not happen for such VVMFs).  Meanwhile $\eta^{-4}$ changes the leading power of $q$ by $q^{-\frac16}$. This happens independently to both of the original $\ell=0$ solutions. So if the original VVMF $\chi^{(0)}$ had leading exponents $(\alpha_0,\alpha_1)$ then the new one $\chi^{(2)}$ has exponents $(\talpha_0,\talpha_1)=(\alpha_0-\frac16,\alpha_1-\frac16)$. Now from the valence formula we have:
\be
\alpha_0+\alpha_1=\frac16
\ee
and hence:
\be
\talpha_0+\talpha_1=-\frac16
\ee
which corresponds to $\ell=2$. As expected, we see that a single action by $\Theta$ augments $\ell$ by 2.

Here we should emphasise an important point. Suppose, as is standard, the starting VVMF has exponents are parametrised by:
\be
\begin{split}
\alpha_0^{(0)} &=-\frac{c^{(0)}}{24}, \qquad
\alpha_1^{(0)} =-\frac{c^{(0)}}{24}+h^{(0)}=\frac{c^{(0)}+4}{24}
\end{split}
\label{alphaell.0}
\ee
where we used the valence formula for $\ell=0$. 
Now the $\Theta$-map shifts all $\alpha_i^{(0)}$ by the same amount $-\frac16$, hence in this parametrisation it only shifts: \be
\begin{split}
c^{(2)} &= c^{(0)}+4, \qquad h^{(2)} = h^{(0)}
\end{split}
\ee
However we are also free to reinterpret the $\chi_i^{(2)}$ in a different way, by exchanging their roles. In this case, we would have:
\be
\begin{split}
\alpha_0^{(2)} &=\frac{c^{(0)}}{24},\qquad 
\alpha_1^{(2)} =-\frac{c^{(0)}+4}{24}
\end{split}
\ee
As a result, with the inverted identification, the $\Theta$-map results in:
\be\label{eq:c2c0}
\begin{split}
c^{(2)} &= -c^{(0)},\qquad h^{(2)} = -h^{(0)}
\end{split}
\ee
Hence in this interpretation, the map cannot relate unitary CFT characters among themselves\,\footnote{Of course we do not know if the characters describe CFT at all, but if $c$ or $h$ is negative then we do know that they cannot describe unitary ones.}. Instead one should think of it acting on the space of quasi-characters, with special cases arising when it acts on a quasi-character to output a (unitary) admissible character or vice versa. Now using Eq.~\eqref{eq:c2c0} we see that:
\be
  c^{(2)}
  =
  -c^{(0)}
  =
  -24M^{(0)}-j^{(0)}
  =
  24(-M^{(0)}-1)+(24-j^{(0)}),
\ee
and so we can identify
\be\label{eq:M2_M0}
  M^{(2)}=-M^{(0)}-1,
  \qquad
  j^{(2)}=24-j^{(0)},
\ee
which is true for the values listed in Eqs.~\eqref{jvalell.0} and \eqref{tjvalell.2}. Therefore, the \(\Theta\)-map sends
\be
  (M^{(0)},j^{(0)})_{\ell=0}
  \to
  (-M^{(0)}-1,\,24-j^{(0)})_{\ell=2},
\ee
with the two characters exchanged.

To illustrate this, let us work out the action of $\Theta$ on some specific MMS solutions.  We start with the $\mA_{1,1}$ WZW model, with $c=1, h=\frac14$. Thus $\alpha_0=-\frac{1}{24},\alpha_1=\frac{5}{24}$. Acting with $\Theta$ we find a pair of characters with exponents $-\frac{5}{24}, \frac{1}{24}$ (the details are provided in Appendix \ref{app:theta_Pn}). This appears to correspond to $c=5, h=\frac14$. But from Eqs.~(\ref{tcqcell.2}),(\ref{tjvalell.2}) we see that there are no $\ell=2$ quasi-characters with this value of $c$. So we are forced to exchange the role of the characters, which gives $c=-1,h=-\frac14$. Now these are a pair of quasi-characters belonging to the dual $\mA_1$ family, with $M^{(2)}=-1, j^{(2)}=23$ in \eref{tcqcell.2}. Similarly one can act with $\Theta$ on the remaining MMS solutions and each time we land on one of the $\ell=2$ quasi-characters.

On the other hand we could start with the $c=-23, h=-\frac74$ quasi-character solution of the MMS equation. Then we end up with what looks like a $c=-19$ solution of the $\ell=2$ MLDE with $h=-\frac74$. Again this does not appear in the list \eref{tcqcell.2}. After exchanging the two solutions we get $c=23, h=\frac74$, which is in the afore-mentioned list and is in fact admissible -- it is a solution originally discovered in \cite{Naculich:1988xv, Hampapura:2015cea} that was subsequently identified with a set of RCFTs in \cite{Gaberdiel:2016zke}. Indeed, it is trivial to reproduce all the rank-2 solutions with $\ell=2$ studied in  the above references starting with suitable $\ell=0$ quasi-characters. 
Concretely, we have verified that all the $\ell=2$ admissible solutions obtained in \cite{Hampapura:2015cea} (see also \cite{Naculich:1988xv}) and later identified as novel coset RCFTs in \cite{Gaberdiel:2016zke} are obtained, bijectively, from $\ell=0$ quasi-character solutions of the MMS equation via the $\Theta$-map.

At this point we pause to resolve a potential puzzle. We saw that acting with $\Theta$ sends the modular $S$-matrix to minus itself. This may seem confusing since it was shown in \cite{Gaberdiel:2016zke} that $\ell=0,2$ solutions satisfy a bilinear relation that pairs them into a modular invariant. For this to be true, the $S$ matrix for the $\ell=2$ solution has to be the inverse of that for the $\ell=0$ solution. Instead we seem to find $S^{(2)}=-S^{(0)}$ which does not work, since $S^2={\cal C}$, the charge conjugation operator, and in the $\mA_1$ family in fact $S^2=1$. The resolution comes from a combination of two effects. One is that when $D$ acts on a solution, it pulls down the leading exponent and if this is negative, the overall sign of one of the solutions changes. Moreover it is well-known that the two exponents $\alpha_0^{(0)}, \alpha_1^{(0)}$ in \eref{alphaell.0} necessarily have opposite signs. Hence $D$ always reverses the overall sign of just one member of the pair. But conventionally, characters and quasi-characters are defined with a positive leading term. So after acting with $\Theta$ we change the overall normalisation back to positive, which flips the sign of only the off-diagonal elements of $S$. The second effect comes from our interchange of the roles of the two solutions, which exchanges the diagonal elements. In view of the general form of the MMS $S$-matrix \cite{Mathur:1988gt}, this just flips the signs of these elements. These two effects together send the $S$-matrix to minus itself, neutralising the negative sign due to the multiplier system of $\eta$. Thus finally we do get the expected $S$-matrix.
 
Next, let us consider repeated actions of $\Theta$, which introduces new features. Iterating and using the fact that the derivative $D$ commutes with $\eta$, we get:
\be
\Theta^n =\eta^{-4n}D^n
\ee
The above arguments then show that this operator augments $\ell$ by $2n$. However we clearly will not find the most general one, since in this process the only free parameter is $\mu$ of the original equation, while the most general MLDEs have an increasing number of independent parameters as $\ell$ increases.

To understand this better, consider the case $n=2$. Then $\Theta^2\chi^{(0)}$ solves the $\ell=4$ equation. However, we can also multiply $\chi^{(0)}$ by $j^\frac13$ to get an $\ell=4$ solution. Hence the most general such solution is:
\be
\chi^{(4)}\equiv(\Theta^2+\beta j^\frac13)\chi^{(0)}
\label{Theta2op}
\ee
where $\beta$ is an arbitrary real parameter. Adding the two terms in the above equation is consistent with the $q$-expansion since each one starts with $q^{-\frac13}$. moreover they are easily verified to have the same $T,S$ transformations.

The solution apparently acquires a new parameter $\beta$.  However in the present example this is actually not the case. We find:
\be\label{red}
\begin{split}
\Theta^2 \chi^{(0)} &= \eta^{-8}D^2\chi^{(0)}\\
& = -\mu\eta^{-8}E_4\chi^{(0)}\\
&= -\mu j^\frac13\chi^{(0)}
\end{split}
\ee
where we used the definition of $j$. Thus the $\ell=4$ solution \eref{Theta2op} just reduces to a multiple of $j^\frac13\chi^{(0)}$ and no new parameter is introduced. This agrees with the results of \cite{Chandra:2018pjq, Das:2023qns} where it is argued that every $\ell=4$ solution takes this form. We thus learn that multiple operations of $\Theta$ can in principle accompanied by a finite set of lower-order operators, apparently introducing new parameters, but that in specific cases these parameters may collapse into a smaller number. Below we look at cases where the new parameters survive.

\subsection{Wronskian index $\ell \ge 6$}

In the above discussion we saw that the $\ell=0,2,4$ solutions all solve a one-parameter MLDE, that determines its $c,h$ values. However once $\ell\ge 6$, genuine additional parameters arise in the MLDE. These are known as {\it non-rigid parameters} since they correspond to movable poles of the MLDE \cite{Das:2023qns}. A new movable pole arises every time $\ell$ jumps by 6. On the other hand the fractional poles are stuck at orbifold points of the torus moduli space and cannot move, hence they are not associated to new parameters. 

We now review some relevant material about $\ell\ge 6$ MLDEs.
For all \(\ell=6r\), the generic rank-two MLDE in \(\tau\)-space is
\begin{equation}
    \left[
        D^2
        +
        E_4^2E_6
        \sum_{I=1}^{r}
        \frac{1}{E_4^3-p_I\Delta}\,D
        +
        \alpha_0^{(6r)}\alpha_1^{(6r)} E_4
        \frac{
            \prod_{I=1}^{r}\left(E_4^3-b_{4,I}\Delta\right)
        }{
            \prod_{I=1}^{r}\left(E_4^3-p_I\Delta\right)
        }
    \right]\chi_i^{(6r)}=0 .
    \label{gen_26r_movable_review}
\end{equation}
Similarly, for \(\ell=6r+2\), one has
\begin{equation}
    \left[
        D^2
        +
        \left(
            \frac{E_6}{3E_4}
            +
            E_4^2E_6
            \sum_{I=1}^{r}
            \frac{1}{E_4^3-p_I\Delta}
        \right)D
        +
        \alpha_0^{(6r+2)}\alpha_1^{(6r+2)} E_4
        \frac{
            \prod_{I=1}^{r}\left(E_4^3-b_{4,I}\Delta\right)
        }{
            \prod_{I=1}^{r}\left(E_4^3-p_I\Delta\right)
        }
    \right]\chi_i^{(6r+2)}=0 ,
    \label{gen_26r2_movable_review}
\end{equation}
where $\Delta(q)\equiv\eta^{24}(q)$ denotes the modular discriminant.

The $\ell=6r+4$ case can be found in \cite{Das:2023qns}. Here, $\alpha_0,\alpha_1$ are the exponents of the two independent solutions, and as before, they carry superscripts to remind us of the $\ell$-value of the equation in which they arise. 

The remaining parameters are $p_I, b_{4,I}, I =1,2,\cdots,r$.
The \(p_I\) are the locations of the movable poles in the MLDE\,\footnote{To be clear, we emphasise that these are not poles of the characters themselves, which are always pole-free. Rather, these poles arise from dividing the equation by a common factor, the Wronskian determinant, which itself has zeroes.}. Thus they are apparent singularities of the differential equation, so the solutions must be single-valued around each of them. The parameters \(b_{4,I}\) appear in the numerator of the weight-four coefficient of the MLDE and are called accessory parameters \cite{Das:2023qns}. They are not fixed by the leading exponents at \(q=0\), but are constrained by regularity at the movable poles. In \cite{Das:2023qns}, the accessory parameters are determined in terms of the pole parameters $p_I$ so that we ultimately have $r$ free parameters. The data $(p_I,b_{4,I})$ discussed above are determined by the generalised Wronskians as given in \eref{Wronskians} via \eref{phidef} (see \cite{Das:2023qns} for more details). We shall see below that the $\Theta$-map procedure \eref{eq:sc_op} will generate the full family of non-rigid MLDEs, and that it provides the movable poles and accessory parameters explicitly.

\paragraph{$(2,0)$ to $(2,6)$ map:} We start with the map from $\ell=0$ to $\ell=6$. For this we have the obvious candidate $\Theta^3$, with leading exponent $q^{-\half}$. Adding all allowed operators depending on $\Theta$ and $j$ that have the same leading exponent (as required by modularity) gives the family of $\ell=6$ solutions:
\be
\chi^{(6)}\equiv\left(\Theta^3+\beta j^{\frac13}\Theta+\gamma(j-1728)^\half\right)\chi^{(0)}, \label{26_mp}
\ee
Each term in this operator shifts the leading exponent by $-\half$. Also under the $S$ transformation, all terms give a minus sign: for the first two terms this follows from \eref{Thetatransf} while for the last term it is a well-known property coming from the fact that $j-1728$ vanishes at the orbifold point $\tau=i$ of torus moduli space. 

Using the MMS equation \eref{2ell0} as well as properties of modular forms, the terms in \eref{26_mp} can be simplified as follows:
\be\label{simp_26_j}
\begin{split}
\Theta^3\chi_i &= \eta^{-12}\left(-\mu E_4D+\frac{\mu}{3}E_6\right)\chi_i^{(0)}, \\[2mm]
\beta \, j^{1/3}\Theta \chi_i &= \beta \, \eta^{-12}E_4D\chi_i^{(0)},\\[2mm]
 \gamma \, (j-1728)^{1/2}\chi_i &= \gamma \, \eta^{-12}E_6\chi_i^{(0)}
\end{split}
\ee
which leads to:
\be
\chi^{(6)}_i = \eta^{-12}\left[(\beta-\mu)E_4D
+\left(\gamma+\frac{\mu}{3}\right)E_6\right]\chi_i^{(0)}
\ee
From this we see that only one new free parameter has been introduced (an overall rescaling of the solution does not change the MLDE). For $\beta\neq \mu$, the free parameter is: $\frac{\gamma+\frac{\mu}{3}}{\beta-\mu}$. Since the original solution $\chi_i^{(0)}$ already depends on one free parameter, this brings the total number to 2 as expected \cite{Chandra:2018pjq}.

To find the MLDE, differentiate $\chi^{(6)}_i$ twice and use  \eref{simp_26_j} to get:
\be
\begin{split}\label{Dpsi_6}
D\chi_i^{(6)} = \, &\eta^{-12}\left[\left(\gamma+\frac{2\mu-\beta}{3}\right)E_6\,D+\left(\mu^2-\beta\mu-\frac{\mu+3\gamma}{6}\right)E_4^2\right]\chi_i^{(0)}, \\
D^2\chi_i^{(6)} = \, & \eta^{-12}\left[\frac{1}{6}\left(\beta-6\gamma+3\mu\left(2\mu-2\beta-1\right)\right)E_4^2D\right.\\ 
&+\left.\frac{1}{9}\left(3\gamma\left(1-3\mu\right) 
+\mu\left(9\beta-12\mu+1\right)\right)E_4E_6\right]\chi_i^{(0)},
\end{split}
\ee
where we have used the Ramanujan identities in \eref{j_ids} 
and the MMS equation.

The above can be summarised as:
\begin{equation}
\label{eq:M6matrix22}
\begin{pmatrix}
\chi^{(6)}_i\\
D\chi^{(6)}_i
\end{pmatrix}
=
\eta^{-12}M_6
\begin{pmatrix}
D\chi_i^{(0)}\\
\chi_i^{(0)}
\end{pmatrix}
\ee
where:
\be
M_6 := 
\begin{pmatrix}
(\beta-\mu)E_4 & \left(\gamma+\frac{\mu}{3}\right)E_6\\[2pt]
\left(\gamma+\frac{2\mu-\beta}{3}\right)E_6 \ \ \ 
&
\left(\mu^2-\beta\mu-\frac{\mu+3\gamma}{6}\right)E_4^2
\end{pmatrix}.
\ee
Now denoting $\chi_0$ and $\chi_1$ as linearly independent solutions of the MMS equation we get the following relations between the generalised Wronskians for the $(2,6)$ and $(2,0)$ MLDEs.
\begin{equation}\label{gen_W_2}
\begin{split}
    &W_2^{(6)} = -\eta^{-24}{\rm det}\,M_6\,W_2^{(0)}, \\
    &W_1^{(6)} = -\eta^{-24}{\rm det}\,N_6\,W_2^{(0)}, \\
    &W_0^{(6)} = -\eta^{-24}{\rm det}\,O_6\,W_2^{(0)},
\end{split}    
\end{equation}
where 
\be
N_6=
\begin{pmatrix}
(\beta-\mu)E_4 & \left(\gamma+\frac{\mu}{3}\right)E_6\\[2pt]
\frac{1}{6}\left(\beta-6\gamma+3\mu\left(2\mu-2\beta-1\right)\right)E_4^2 \ \
&
\frac{1}{9}\left(3\gamma\left(1-3\mu\right)+\mu\left(9\beta-12\mu+1\right)\right)E_4E_6
\end{pmatrix}
\ee
and 
\be
O_6=
\begin{pmatrix}
\left(\gamma+\frac{2\mu-\beta}{3}\right)E_6 \ \
&
\left(\mu^2-\beta\mu-\frac{\mu+3\gamma}{6}\right)E_4^2\\[2pt]
\frac{1}{6}\left(\beta-6\gamma+3\mu\left(2\mu-2\beta-1\right)\right)E_4^2 \ \
&
\frac{1}{9}\left(3\gamma\left(1-3\mu\right)+\mu\left(9\beta-12\mu+1\right)\right)E_4E_6
\end{pmatrix}
\ee

Now the generic  MLDE for $\ell=6$ is found by specialising \eref{gen_26r_movable_review} to $r=1$ (it is written explicitly in Eq.~(2.21) of \cite{Das:2023qns}), and has the coefficient functions:
\begin{equation}\label{phis_26}
    \phi_2 = \frac{E_4^2 E_6}{E_4^3-p\Delta}, \qquad \phi_4 = \alpha_0^{(6)}\alpha_1^{(6)}E_4\frac{E_4^3-b_{4,1}\Delta}{E_4^3-p\Delta}    
\end{equation}
where $\alpha_0^{(6)}\alpha_1^{(6)} \equiv \mu+\frac{1}{6} = \left(\alpha_0^{(0)}-\frac{1}{2}\right)\left(\alpha_1^{(0)}-\frac{1}{2}\right)$. Moreover the values of $p,b_{4,1}$ are found to be:
\begin{equation}
\label{eq:pbetagamma}
p = 1728\, \frac{
\left(\gamma+\frac{\mu}{3}\right)
\left[
\gamma+\frac{\mu}{3}
-\frac13(\beta-\mu)
\right]
}{
\mu(\beta-\mu)^2
+\frac16(\beta-\mu)\left(\gamma+\frac{\mu}{3}\right)
+
\left(\gamma+\frac{\mu}{3}\right)^2
},
\end{equation}
and
\begin{equation}
\label{eq:bbetagamma}
b_{4,1}
=
\frac{
192\left[
(\beta-\mu)-3\left(\gamma+\frac{\mu}{3}\right)
\right]
\left[
3\mu\left((\beta-\mu)-\left(\gamma+\frac{\mu}{3}\right)\right)
+\left(\gamma+\frac{\mu}{3}\right)
\right]
}{
\left(\mu+\frac16\right)
\left[
\mu(\beta-\mu)^2
+\frac16(\beta-\mu)\left(\gamma+\frac{\mu}{3}\right)
+
\left(\gamma+\frac{\mu}{3}\right)^2
\right]
}.
\end{equation}
Hence, $\chi^{(6)}_i$ as obtained from \eref{26_mp} using our $\Theta$-map indeed solves a generic $\ell=6$ MLDE. Following similar arguments, the generic map to get from the solutions of the $(2,0)$ MLDE to those for $(2,6r)$ can be constructed and we leave this as an exercise to the reader. 

Most generally, we consider polynomial operators of the form:
\be
\sum_{2p+3q+r=n} X_{pqr}\, j^{\frac{p}{3}}(j-1728)^{\frac{q}{2}}\,\Theta^{r}
\label{genThetaop}
\ee
where $X_{pqr}$ are arbitrary coefficients. It is easily verified that all terms in the polynomial acquire the same factor $(-1)^{2p+3q+r}$ under the modular $S$-transformation, as well as the same phase under $T$. Thus this operator maps VVMFs to VVMFs and raises $\ell$ by $2n$. Starting from a $\ell=0$ solution, we find an $\ell=2n$ solution with the same number of free parameters as the most general equation of this type. 

A key output of our approach is the preservation of integrality. The approach starts with quasi-character at $\ell=0$, which is already integral in the sense noted earlier,  and uses the differential operator $D$ as well as the modular ingredients $\eta, E_2, j^\frac13, (j-1728)^\half$. Differentiation preserves integrality (possibly after a change of overall normalisation) and the other ingredients all have integral (though not positive) $q$-series expansions. It only remains to take the coefficients $X_{abc}$ to be integers or at least rational numbers. As a result, the new coefficients remain rational with bounded denominators and therefore can be rendered integral with a suitable choice of normalisation.

Since all allowed values of $\ell$ are even for rank-2 VVMFs, we conclude that {\em differential operators of the form \eref{genThetaop} generate all rank-2 quasi-characters starting from the quasi-character solutions of the MMS equation}.

We pointed out in some examples that in order to relate quasi-characters using $\Theta$, it is necessary to exchange the identification of the components $\chi_0,\chi_1$ after carrying out the $\Theta$ action. Let us now see how this works when $\Theta$ is replaced by $\Theta^n$. One easily finds:
\be\label{eq:ex_al_2n}
\begin{split}
\alpha_0^{(2n)} & = \frac{c^{(0)}+4(1-n)}{24},\qquad
\alpha_1^{(2n)} = -\frac{c^{(0)}+4n}{24}
\end{split}
\ee
which means:
\be
\begin{split}
c^{(2n)} &= -c^{(0)} - 4(1-n),\qquad
h^{(2n)} = -h^{(0)}
\end{split}
\ee

\section{Relating higher-rank MLDEs by the $\Theta$-map}
\label{app:rank_three}

In rank 2, we saw that every action of the $\Theta$-map augments the Wronskian index by 2. We now show that in rank $p$, the same map augments $\ell$ by $p$. This follows from the valence formula. Suppose we start with a set of exponents $\alpha_i$ that satisfy \eref{valence} for some given $p$ and $\ell$. The $\Theta$-map shifts each $\alpha_i$ by $-\frac16$, so the LHS of the equation changes to:
\be
\begin{split}
\sum_{i=0}^{p-1}\left(\alpha_i-\frac16\right) &= \sum_{i=0}^{p-1}\alpha_i - \frac{p}{6}\\[2mm]
&= \frac{p(p-1)}{12}-\frac{\ell+p}{6}
\end{split}
\ee
and we see that $\ell$ has shifted by $p$.

We already saw how this works in rank 2 -- the shift of $\ell$ is by 2, and since $\ell$ is known to be even, this means we generate all possible quasi-character VVMFs just starting from $\ell=0$. But there, the general solution for all $\ell$ was already known from the linear combination method described in \cite{Chandra:2018pjq} so despite the convenience of the new method, it did not generate significant new information for us. However in rank $3$ the situation is more interesting, as it  was shown in \cite{Kaidi:2021ent} that $\ell$ must be a multiple of 3. This, along with the invertibility of the map, immediately tells us that the $\Theta$-map will generate {\em all} VVMFs of quasi-character type starting from those with $\ell=0$. Now, very little is known about quasi-characters for rank 3 and $\ell>0$. For $\ell=0$ there is a partial classification \cite{Mukhi:2020gnj, Govindarajan:2025rgh, Govindarajan:2025jlq} with the last of these papers also containing some $\ell=3$ examples. Additional quasi-characters with $\ell=0$ have recently been found \cite{DasMukhi:unpub}. It seems quite feasible to find a complete classification of quasi-characters for rank 3 and $\ell=0$, and in view of the above discussion this would enable us to find all quasi-character solutions for all $\ell$.

\subsection{Rank 3}\label{sec:r3}

In this sub-section we discuss the $\Theta$-map on rank $3$ MLDEs and their images under the $\Theta$-map, while in the following sub-section we briefly comment on higher rank MLDEs.

Let
\(\chi_i^{(0)}\), \(i=0,1,2\), be three linearly independent solutions of the
\(\ell=0\) MLDE
\begin{equation}
    \left(
        D^3
        +
        \mu_1 E_4 D
        +
        \mu_2 E_6
    \right)\chi_i^{(0)}=0 .
    \label{rank3_30_mlde}
\end{equation}
Labelling the solutions as:
\be
    \chi_i^{(0)}(q)=q^{\alpha_i^{(0)}}\big(1+O(q)\big),
\ee
the exponents are related by the valence formula \eref{valence} which gives:
\be
 \qquad
    \alpha_0^{(0)}+\alpha_1^{(0)}+\alpha_2^{(0)}=\frac12
\ee

As before the $\Theta$-map shifts each non-zero leading exponent by $-\frac16$. Thus the new exponents satisfy:
\be
    \alpha_i^{(3)}
    =
    \alpha_i^{(0)}-\frac16,
    \qquad
    \alpha_0^{(3)}+\alpha_1^{(3)}+\alpha_2^{(3)}=0 .
    \label{rank3_exponent_shift}
\end{equation}
This precisely satisfies the valence formula for a \((3,3)\) MLDE. Equivalently, the map sends
\be
    (c,h_1,h_2)\to (c+4,h_1,h_2),
\ee
Of course we are still free to carry out reordering of the resulting characters. This is more complicated than in the rank-2 case because we have two choices of the new identity character, namely the one originally assigned to $h_1$ or $h_2$. However it is straightforward to work out the resulting formulae in any specific case.

Let us now define $\chi_i^{(3)}=\Theta \chi_i^{(0)}$. Then \eref{rank3_30_mlde} gives:
\begin{equation}
    \mu_2 E_6\chi_i^{(0)}
    =
    -\eta^4
    \left(
        D^2
        +
        \mu_1 E_4
    \right)\chi_i^{(3)} .
    \label{rank3_eliminate_seed}
\end{equation}
Next, let us act with \(D\) on the various terms of Eq.~\eqref{rank3_30_mlde} and re-use $D\chi_i^{(0)} = \eta^4\chi_i^{(3)}$ to get:
\begin{equation}
\begin{split}
    &D^4\chi_i^{(0)}
    =
    \eta^4D^3\chi_i^{(3)},\\
    &D(E_4D\chi_i^{(0)})
    =
    D(E_4\eta^4\chi_i^{(3)})
     =
    \eta^4
    \left(
        E_4D-\frac13E_6
    \right)\chi_i^{(3)},\\
    &D(E_6\chi_i^{(0)})
    =
    E_6D\chi_i^{(0)}
    -
    \frac12E_4^2\chi_i^{(0)}
     =
    \eta^4E_6\chi_i^{(3)}
    -
    \frac12E_4^2\chi_i^{(0)} .
\end{split}    
\end{equation}
Therefore
\begin{equation}
    \eta^4
    \left[
        D^3
        +
        \mu_1 E_4D
        +
        \left(\mu_2-\frac{\mu_1}{3}\right)E_6
    \right]\chi_i^{(3)}
    -
    \frac{\mu_2}{2}E_4^2\chi_i^{(0)}
    =
    0 .
\end{equation}
Eliminating \(\chi_i^{(0)}\) using Eq.~\eqref{rank3_eliminate_seed}, we find
\begin{equation}
    \left[
        D^3
        +
        \frac12\frac{E_4^2}{E_6}D^2
        +
        \mu_1 E_4D
        +
        \left(\mu_2-\frac{\mu_1}{3}\right)E_6
        +
        \frac{\mu_1}{2}\frac{E_4^3}{E_6}
    \right]\chi_i^{(3)}
    =
    0 .
\end{equation}
Using \(E_4^3-E_6^2=1728\Delta\), the above becomes
\begin{equation}
    \left[
        D^3
        +
        \frac12\frac{E_4^2}{E_6}D^2
        +
        \mu_1 E_4D
        +
        \left(\mu_2+\frac{\mu_1}{6}\right)E_6
        +
        864\mu_1\,\frac{\Delta}{E_6}
    \right]\chi_i^{(3)}
    =
    0,
    \label{rank3_33_image}
\end{equation}
Let us now compare the above with the general \(\ell=3\) equation as was studied in \cite{Gowdigere:2023xnm}:
\begin{equation}
    \left[
        D^3
        +
        \frac12\frac{E_4^2}{E_6}D^2
        +
        \nu_1 E_4D
        +
        {\nu}_2 E_6
        +
        {\nu}_3\,\frac{\Delta}{E_6}
    \right]\chi_i^{(3)}
    =
    0 .
    \label{rank3_33_raw}
\end{equation}
This appears to be a three-parameter
family. However, imposing local regularity of solutions at \(\tau=i\) removes one parameter. Indeed, the local exponents at \(\tau=i\) are: $\{0,\frac12,\frac32\}$. Now since the last two differ by an integer, a logarithmic solution can appear. Requiring the absence of logarithmic terms gives (see Appendix \ref{app:loc_reg} for more details):
\begin{equation}
    {\nu}_3=864\,\nu_1 .
    \label{rank3_nolog_condition}
\end{equation}
Thus the \(\ell=3\) family actually has only two free parameters:
\(\nu_1\) and \(\nu_2\)\,\footnote{This explains the fact, noted in Eq. (58) of \cite{Gowdigere:2023xnm}, that all admissible solutions satisfy the above relation.}. Eq.~\eqref{rank3_33_image} manifestly satisfies the above regularity condition, with parameters:
\begin{equation}
\nu_1=\mu_1,\qquad
    \nu_2=\mu_2+\frac{\mu_1}{6},
    \qquad
    \nu_3=864\mu_1 .
    \label{rank3_image_parameters}
\end{equation}
Hence the $\Theta$-map produces the full \(\ell=3\) family of solutions.

The inverse map is straightforward. From Eq.~\eqref{rank3_eliminate_seed}, whenever
\(\mu_2\neq0\),
\begin{equation}
    \chi_i^{(0)}
    =
    -\frac{\eta^4}{\mu_2 E_6}
    \left(
        D^2
        +
        \mu_1 E_4
    \right)\chi_i^{(3)} ,
    \label{rank3_inverse_AB}
\end{equation}
where, as was shown above, $\chi_i^{(3)}$ indeed satisfies Eq.~\eqref{rank3_33_image}. Thus, away from the case $\mu_2=0$, we have a one-to-one map between $\ell=0$ and $\ell=3$ quasi-characters\,\footnote{If \(\mu_2=0\), it is easy to see that the differential operator in \eref{rank3_30_mlde} factorises into $(D^2+\mu_1 E_4)D$ and hence essentially reduces to the MMS operator with additional constant solutions.}.

As indicated above, we have restricted our attention to non-logarithmic solutions. It may be interesting to explore the logarithmic case in the present context, as such characters frequently arise in the 4d-2d correspondence \cite{Beem:2017ooy}.

Repeated application of the $\Theta$-map together with suitable functions of $j$, as in \eref{genThetaop}, can be seen to generate MLDE solutions for all $\ell=3r$ where $r$ is a positive integer. Remarkably it has been shown \cite{Kaidi:2021ent} that all well-behaved rank-3 MLDEs have $\ell=3r$ and therefore our $\Theta$-map will generate all rank-3 quasi-characters if one only knows the complete set with $\ell=0$. As mentioned above, a partial list of the latter was provided in \cite{Mukhi:2020gnj} and more have been discovered since then \cite{Gowdigere:2023xnm, Rayhaun:2023pgc, Govindarajan:2025rgh, Govindarajan:2025jlq, DasMukhi:unpub} though a complete classification does not exist so far.

\subsubsection{An $\ell=3$ example and its pre-image under the $\Theta$-map}

A recently discovered $\ell=3$ RCFT can be found in \cite{Rayhaun:2023pgc} see the discussion around Eq.~(201). It is denoted $\mathcal{E}_3[\mA_{1,5}\otimes \mE_{7,1}]$ which means it is a three-character extension of a (twelve-primary) WZW tensored-product RCFT: $\mA_{1,5}\otimes \mE_{7,1}$. This theory has $(c^{(3)},h_1^{(3)},h_2^{(3)}) = (\frac{64}{7},\frac67,\frac27)$ and using Eq.~\eqref{valence} one can readily see that it is an $\ell=3$ solution. In the classification of low-rank modular tensor categories \cite{rowell2009classification}, it is associated to the $(\mA_1,5)_{\frac12}$ MTC.

Now from the inverse $\Theta$-map construction above, we can readily find the $\ell=0$ solution that is mapped by $\Theta$ to the above RCFT. A simple computation, using Eq.~\eqref{rank3_exponent_shift}, shows that it should have  $(c^{(0)}, h_1^{(0)}, h_2^{(0)}) = (\frac{36}{7}, \frac67, \frac27)$. This solution is a quasi-character, as can be seen from its explicit $q$-series below:
\begin{equation}
\begin{split}
  \chi_0^{(0)}
  &=
  q^{-3/14}
  \left(
    1 - 36\,q - 225\,q^2 - 1164\,q^3 - 4599\,q^4
    +\ldots
  \right),
  \\[2mm]
  \chi_1^{(0)}
  &=
  q^{9/14}
  \left(
    39 + 324\,q + 1674\,q^2 + 6920\,q^3 + 24174\,q^4
    +\ldots
  \right),
  \\[2mm]
  \chi_2^{(0)}
  &=
  q^{1/14}
  \left(
    9 + 124\,q + 855\,q^2 + 3924\,q^3 + 15000\,q^4
    +\cdots
  \right).
\end{split}  
\end{equation}
This solution, with $c=\frac{36}{7}$, nicely confirms our general picture. It has previously appeared, in a different context, in Table 12, row 5 of \cite{Duan:2022ltz}. It can also be found using Bantay-Gannon theory, by starting with the characteristic matrix in rows 5 and 10 of Table 3 of \cite{Govindarajan:2025jlq} and applying a Bantay-Gannon recursion relation to the middle column \footnote{We thank Suresh Govindarajan for bringing these references to our attention.}.

\subsection{Rank $4$ and above}

A special feature of ranks 2 and 3 is that the quasi-character solutions to all MLDEs are generated by repeated application of our $\Theta$-map in the form of \eref{genThetaop} to the $\ell=0$ solutions. This simplifies enormously the search for admissible VVMFs with two and three characters, allowing us to go far beyond the small number of cases worked out so far. Unfortunately this feature fails to hold from rank 4 onwards. We have seen that $\Theta$ sends $\ell\to \ell+p$. However there is no restriction, as far as we know, that the $\ell$-value of a rank-$p$ VVMF must be a multiple of $p$. In fact, it was shown in \cite{Kaidi:2021ent} that for rank 4, $\ell$ is even and for rank 5, any non-negative integral $\ell\neq 1$ is allowed. 

Examples of rank-4 RCFTs solving an $\ell=2$ MLDE can be found in Table 7 of \cite{Govindarajan:2026cgv}. Table 9 of \cite{Govindarajan:2026cgv} similarly lists admissible rank-5 solutions solving an $\ell=2$ MLDE, which have been identified as Hecke images of various minimal models in \cite{Duan:2022ltz} (see \cite{Harvey:2018rdc} for more details on Hecke images and RCFTs). Clearly these cannot be generated by the $\Theta$-map from $\ell=0$ solutions in rank 4 or 5. However, the map still brings about a major simplification: for any rank $p$, if we know all the quasi-character solutions with $\ell\in\{0,2,3,\ldots,p-1,p+1\}$, the $\Theta$-map does the rest of the job by generating all solutions with $\ell > p+1$.

\section{Application of the $\Theta$-map: signs of $\ell=2$ quasi-characters}\label{sec:l2-signs-theta}

In this section, we will use the $\Theta$-map to prove, for the first time, a conjecture \cite{Chandra:2018pjq, Das:2025gto} regarding the alternation of signs in the coefficients of $\ell=2$ quasi-character solutions in rank-2. These sign patterns and additionally the rate at which these $q$-series coefficients of the above quasi-characters grow will be extremely useful when forming linear combinations of them to get admissible character-like solutions to $\ell\geq 6$ MLDE in rank-2 \cite{Chandra:2018pjq, Das:2025gto}. For the $\ell=0$ quasi-characters a geometric growth was proven in \cite{Das:2025gto} for their $q$-series coefficients\footnote{It was found in \cite{Das:2025gto} that \(\lvert a_n\rvert\ge R\lvert a_{n-1}\rvert\), with $R>1$, for $n\leq 2|M^{(0)}|$. The $R=1$ growth happens when $n\gg |c|$. In this case, we enter the Rademacher regime where the necessary condition for Rademacher behaviour to kick in is: $\left|\frac{a_{n+1}}{a_n}\right|=R\simeq 1$ \cite{Das:2025gto, Chandra:2018pjq}.}. We will be able to show an analogous geometric growth for the $\ell=2$ coefficients also. We first review the conjecture for both $\ell=0$ and $\ell=2$ families. Thereafter we will use the $\Theta$-map to prove the latter starting with our recent proof of the former in \cite{Das:2025gto}.

We start with $\ell=0$. The central charge is a function of $(M^{(0)},j^{(0)})$ as given in Eq.~\eqref{cqcell.0}, and hence the $q$-series coefficients $a_{i,n}^{(0)}$ depend on these two parameters. It was observed in \cite{Chandra:2018pjq} and proven in \cite{Das:2025gto} that these coefficients obey the following sign pattern. For $M^{(0)}>0$ and for $i=0$ (identity component), all odd coefficients are negative while all even coefficients are positive in the range $0\leq n\leq 2M^{(0)}$. Beyond this, all coefficients are positive. For $i=1$ (non-identity component), all coefficients are positive. For $M^{(0)}<0$ the roles are roughly exchanged: for $i=0$ all coefficients are positive while for $i=1$ all odd coefficients are negative and all even coefficients are positive for $0\leq n\leq 2|M^{(0)}|-1$. Beyond this, all coefficients are negative. Recall that for $M^{(0)}=0$, we have the admissible MMS solutions where all coefficients of both components are positive.

A different sign pattern was observed for the $\ell=2$ quasi-characters in \cite{Chandra:2018pjq}. Here the $q$-series coefficients $a_{i,n}^{(2)}$ are functions of $(M^{(2)},j^{(2)})$ as in Eq.~\eqref{tcqcell.2}. For $M^{(2)}>0$ and for $i=0$ all odd coefficients are negative while all even coefficients are positive for $0\leq n\leq M^{(2)}$. After this the sign alternation pattern again reverses -- all even coefficients turn negative while all odd coefficients are positive, in the range $M^{(2)}+1\leq n\leq 2M^{(2)}$. Beyond this, all coefficients are positive. For $i=1$, all coefficients are positive. For $M^{(2)}<0$ again the roles are again (roughly) exchanged: for $i=0$ all coefficients are positive while for $i=1$, all odd coefficients are negative and all even coefficients are positive for $0\leq n\leq |M^{(2)}|-1$. After this the sign alternation pattern again reverses -- all even coefficients turn negative while all odd coefficients turn positive for $|M^{(2)}|\leq n\leq 2|M^{(2)}|-2$. Again, for $M^{(2)}=0$ we obtain admissible solutions, this time the ones found in \cite{Naculich:1988xv, Hampapura:2015cea, Gaberdiel:2016zke}.

In \cite{Das:2025gto}, the sign patterns for the $\ell=0$ family were proven by making use of the Frobenius recursion relation for the MMS equation Eq.~\eqref{2ell0}. However the recursion relation needed to solve the $\ell=2$ equation Eq.~\eqref{2ell2} is more involved and so in \cite{Das:2025gto} we were unable to prove the $\ell=2$ sign pattern. However we have now proved it (see below) using the $\Theta$-map to relate $\ell=0$ to $\ell=2$.

For simplicity, for $\ell=0,2$, we will focus on the $\mA_1$ and dual $\mA_1$ families of quasi-characters, obtained by taking $j^{(0)}=1$ in Eq.~\eqref{jvalell.0} and $j^{(2)}=23$ in Eq.~\eqref{tjvalell.2} respectively. The sign patterns for other families of $\ell=2$ quasi-characters can be readily proven using similar arguments.

The \(\mA_1\) family of quasi-characters can be expressed as:
\begin{equation}
\chi^{(0)}_{0,M^{(0)}}(q)
=
q^{-M^{(0)}-\frac{1}{24}}
\sum_{n\ge0}a^{(0)}_{0,n}(M^{(0)})q^n,
\qquad
\chi^{(0)}_{1,M^{(0)}}(q)
=
q^{M^{(0)}+\frac{5}{24}}
\sum_{n\ge0}a^{(0)}_{1,n}(M^{(0)})q^n,
\label{eq:l0-A1-def}
\end{equation}
with the leading coefficients normalised to be positive, while the dual \(\mA_1\) family of quasi-characters is:
\begin{equation}
\chi^{(2)}_{0,M^{(2)}}(q)
=
q^{-M^{(2)}-\frac{23}{24}}
\sum_{n\ge0}a^{(2)}_{0,n}(M^{(2)})q^n,
\qquad
\chi^{(2)}_{1,M^{(2)}}(q)
=
q^{M^{(2)}+\frac{19}{24}}
\sum_{n\ge0}a^{(2)}_{1,n}(M^{(2)})q^n,
\label{eq:l2-A1-def}
\end{equation}
again with positive leading coefficients.

\subsection{Sign patterns for \(\ell=0\) quasi-characters}

To begin, we summarise the results for the signs and growth of coefficients for the \(\ell=0\) \(\mA_1\)-family of quasi-characters that were derived in \cite{Das:2025gto}:
\paragraph{} For \(M^{(0)}>0\):
\begin{itemize}
\item[i)] For the identity character, the coefficients satisfy
\be\label{eq:N>0_id_0}
        a^{(0)}_{0,n}(M^{(0)})
        =
        (-1)^n|a^{(0)}_{0,n}(M^{(0)})|
        \qquad(0\le n\le2M^{(0)}),
\ee
and
\be\label{eq:N>0_id_1}
        a^{(0)}_{0,n}(M^{(0)})>0
        \qquad(n\ge2M^{(0)}+1).
\ee
and we have the following lower bound on the growth \footnote{For the analysis of the $M^{(2)}<0$ case below, we shall use an $\mA_1$-specific strengthening of this estimate with $96$ in Eq.~\eqref{eq:N>0_id_2} replaced by $200$. This stronger estimate is not part of the family-uniform result
of \cite{Das:2025gto}; it is proven in appendix~\ref{app:A1-growth-200}.},
\be\label{eq:N>0_id_2}
        |a^{(0)}_{0,n}(M^{(0)})|
        >
        96|a^{(0)}_{0,n-1}(M^{(0)})|
        \qquad(1\le n\le2M^{(0)})
\ee
For the non-identity character, the coefficients satisfy:
\be\label{eq:N>0_nid_0}
        a^{(0)}_{1,n}(M^{(0)})>0
        \qquad (\hbox{all }n)
\ee
and the growth is bounded as follows:
\be\label{eq:N>0_nid_1}
        a^{(0)}_{1,n}(M^{(0)})
        \ge
        2\,a^{(0)}_{1,n-1}(M^{(0)})
        \qquad(1\le n\le2M^{(0)}).
\ee
\end{itemize}

\paragraph{} For \(M^{(0)}<0\):
\begin{itemize}
\item[ii)] the identity coefficients satisfy
\be\label{eq:N<0_id_0}
        a^{(0)}_{0,n}(M^{(0)})>0
        \qquad(\text{all }n)
\ee
and the growth is now bounded as:
\be\label{eq:N<0_id_1}
        a^{(0)}_{0,n}(M^{(0)})
        \ge
        4\,a^{(0)}_{0,n-1}(M^{(0)})
        \qquad(1\le n\le2|M^{(0)}|)
\ee
while the non-identity coefficients satisfy
\be\label{eq:N<0_nid_0}
        a^{(0)}_{1,n}(M^{(0)})
        =
        (-1)^n|a^{(0)}_{1,n}(M^{(0)})|
        \qquad(0\le n\le2|M^{(0)}|-1),
\ee
and
\be\label{eq:N<0_nid_1}
        a^{(0)}_{1,n}(M^{(0)})<0
        \qquad(n\ge2|M^{(0)}|).
\ee
Moreover, we also have
\be\label{eq:N<0_nid_2}
        |a^{(0)}_{1,n}(M^{(0)})|
        >
        45|a^{(0)}_{1,n-1}(M^{(0)})|
        \qquad(1\le n\le2|M^{(0)}|-1).
\ee
\end{itemize}

\subsection{Main result}

\subsubsection{Action of the \(\Theta\)-map}

Using Eq.~\eqref{eq:M2_M0} we have derived a relation between a given solution with $\ell=0$ and the corresponding $\ell=2$ solution after exchange of components (for the $\mA_1$ family). Parametrising $c^{(0)}=24M^{(0)}+1$, $c^{(2)}=24M^{(2)}+23$, we found:
\be
M^{(0)}=-M^{(2)}-1
\label{M0M2}
\ee
The relations between the solutions are then, for \(M^{(2)}>0\):
\begin{equation}
        \chi^{(2)}_{0,M^{(2)}}
        =
        -\Theta\chi^{(0)}_{1,-M^{(2)}-1},
        \qquad
        \chi^{(2)}_{1,M^{(2)}}
        =
        \Theta\chi^{(0)}_{0,-M^{(2)}-1}.
\label{eq:Theta-M-positive}
\end{equation}
and for \(M^{(2)}<0\),
\begin{equation}
        \chi^{(2)}_{0,-|M^{(2)}|}
        =
        \Theta\chi^{(0)}_{1,|M^{(2)}|-1},
        \qquad
        \chi^{(2)}_{1,-|M^{(2)}|}
        =
        -\Theta\chi^{(0)}_{0,|M^{(2)}|-1}.
\label{eq:Theta-M-negative}
\end{equation}
The overall signs were fixed by demanding positive leading coefficients.

\subsubsection{Identity coefficients for \(M^{(2)}>0\)}

Substituting \eref{M0M2} in Eqs. (\ref{eq:N<0_nid_0})-(\ref{eq:N<0_nid_2}) we get:
\be\label{eq:a1ns}
\begin{split}
a^{(0)}_{1,n}(-M^{(2)}-1)
&=(-1)^n |a^{(0)}_{1,n}(-M^{(2)}-1)|,
\qquad
0\le n\le2M^{(2)}+1,\\
a^{(0)}_{1,n}(-M^{(2)}-1)&<0
\qquad\qquad\qquad\qquad\qquad\quad\quad n\ge 2M^{(2)}+2, \\
|a^{(0)}_{1,n}(-M^{(2)}-1)|&>45|a^{(0)}_{1,n-1}(-M^{(2)}-1)|,
\qquad\,\,\, 1\le n\le2M^{(2)}+1.
\end{split}
\ee

Using the first equation in Eq.~\eqref{eq:Theta-M-positive}, under the action of the $\Theta$-map we get for the $\ell=2$ identity coefficients:
\begin{equation}
        a^{(2)}_{0,n}(M^{(2)})
        =
        -\sum_{s=0}^{n}E_s\left(
        n-s-M^{(2)}-\frac{19}{24}
        \right)
        \left(a^{(0)}_{1,n-s}(-M^{(2)}-1)\right)\equiv \sum_{s=0}^{n}E_s\,d_{n-s},
\label{eq:Mpos-id-conv}
\end{equation}
where $E_s$ is defined as follows:
\begin{equation}
        \eta^{-4}
        =
        q^{-1/6}E(q)
        \equiv q^{-1/6}
        \sum_{s\ge0}E_s\,q^s,
\label{eq:E-kernel-def}
\end{equation}
with $E_0=1$ and $E_s>0\,\,\text{for }s\ge1$, while $d_n$ is defined as follows:
\begin{equation}
        d_n
        \equiv
        -
        \left(
        n-M^{(2)}-\frac{19}{24}
        \right)
        \left(a^{(0)}_{1,n}(-M^{(2)}-1)\right).
\label{eq:Mpos-id-raw}
\end{equation}
The minus sign in Eq.~\eqref{eq:Mpos-id-raw} is precisely the sign in the first equation of Eq.~\eqref{eq:Theta-M-positive}. In particular,
\be
        d_0
        =
        \left(M^{(2)}+\frac{19}{24}\right)\left(a^{(0)}_{1,0}(-M^{(2)}-1)\right)>0,
\ee
so the leading $\ell=2$ identity coefficient is positive. The pre-factor $-\left(n-M^{(2)}-\frac{19}{24}\right)$ in Eq.~\eqref{eq:Mpos-id-raw} is positive for \(0\le n\le M^{(2)}\) and negative for \(n\ge M^{(2)}+1\). Therefore:
\be
        \operatorname{sgn}(d_n)=(-1)^n,\quad 0\le n\le M^{(2)}; \qquad
        \operatorname{sgn}(d_n)=(-1)^{n+1},
        \quad M^{(2)}+1\le n\le2M^{(2)}+1.
\ee
This is exactly the sign pattern that we are trying to predict for the coefficients of the $\ell=2$ solutions. 
So if we can show $\text{sgn}(a_{0,n}^{(2)}(M^{(2)}))=\text{sgn}(d_n)$ then we are done. We do this in appendix \ref{app:sgn_an_dn} where we also prove that for $n\geq 2M^{(2)}+1$ all coefficients are positive. Thus, we have:
\begin{equation}
\operatorname{sgn}\left(a^{(2)}_{0,n}(M^{(2)})\right)
=
\begin{cases}
(-1)^n, & 1\le n\le M^{(2)},\\
(-1)^{n+1}, & M^{(2)}+1\le n\le2M^{(2)},\\
+1, & n\ge2M^{(2)}+1.
\end{cases}
\label{eq:Mpos-id-sign-final}
\end{equation}

In appendix \ref{app:sgn_an_dn} we further prove the following lower bound on the geometric growth of the coefficients:
\begin{equation}
        \left|a^{(2)}_{0,n}(M^{(2)})\right|
        >
        2\left|a^{(2)}_{0,n-1}(M^{(2)})\right|
        \qquad(1\le n\le2M^{(2)}+1).
\label{eq:Mpos-id-supergrowth}
\end{equation}

\subsubsection{Non-identity coefficients for \(M^{(2)}>0\)}

Substituting Eq.~\eqref{M0M2} in Eq.~\eqref{eq:N<0_id_0}, we get:
\be 
        a^{(0)}_{0,n}(-M^{(2)}-1)>0,
        \quad  n\ge0.
\ee
Using the second equation in Eq.~\eqref{eq:Theta-M-positive}, we can express the non-identity $\ell=2$ coefficients as:
\begin{equation}
        a^{(2)}_{1,n}(M^{(2)}) = \sum_{s=0}^n E_s \bigg(n-s+M^{(2)}+\frac{23}{24}\bigg)\left(a^{(0)}_{0,n-s}(-M^{(2)}-1)\right)
        \equiv
        \sum_{s=0}^{n}
        E_s\, e_{n-s},
\label{eq:Mpos-nonid-conv}
\end{equation}
where $E_s$ are as before, and:
\be\label{eq:defn_el}
    e_n\equiv \bigg(n+M^{(2)}+\frac{23}{24}\bigg)\left(a^{(0)}_{0,n}(-M^{(2)}-1)\right).
\ee
Note that every factor in every summand above in Eq.~\eqref{eq:Mpos-nonid-conv} is positive. This is because
\be
        E_s>0,\qquad
        n-s+M^{(2)}+\frac{23}{24}>0,\qquad
        a^{(0)}_{0,n-s}(-M^{(2)}-1)>0.
\ee
Hence, $a^{(2)}_{1,n}(M^{(2)})>0 \,\,(\text{for }n\ge0)$ and we are done. In appendix \ref{app:sgn_an_dn} we also prove the following lower bound on the geometric growth of the non-identity coefficients:
\be\label{eq:Mpos-nonid-supergrowth}
        a^{(2)}_{1,n}(M^{(2)})
        \ge
        4\,a^{(2)}_{1,n-1}(M^{(2)}),
        \quad 1\le n\le2M^{(2)}+2.
\ee

Using similar arguments as above -- specifically, the two relations in Eq.~\eqref{eq:Theta-M-negative},
together with the convolution estimates and the $\mA_1$-specific
strengthening proved in appendix~\ref{app:A1-growth-200}, one obtains
the corresponding results for the case with $M^{(2)}<0$. The only additional
subtlety is a non-uniform growth ratio at the crossover of the two
alternating regions -- which is given in
appendix~\ref{app:Mneg-proof}. The identity coefficients satisfy
\be\label{eq:Mneg-id-positive}
 a^{(2)}_{0,n}\!\left(M^{(2)}\right)>0,
 \qquad n\geq0,
\ee
while the non-identity coefficients obey
\be\label{eq:Mneg-nonid-sign-final}
\operatorname{sgn}\left(a^{(2)}_{1,n}\!\left(M^{(2)}\right)\right)
=
\begin{cases}
(-1)^n,
&0\leq n\leq |M^{(2)}|-1,\\
(-1)^{n+1},
&|M^{(2)}|\leq n\leq2|M^{(2)}|-2,\\
-1,
&n\geq2|M^{(2)}|-1.
\end{cases}
\ee
Moreover, for $|M^{(2)}|\geq2$,
\be\label{eq:Mneg-nonid-growth}
 \left|a^{(2)}_{1,n}\!\left(M^{(2)}\right)\right|
 >
 2
 \left|a^{(2)}_{1,n-1}\!\left(M^{(2)}\right)\right|,
 \qquad
 1\leq n\leq2|M^{(2)}|-2.
\ee

Thus we see that the $\Theta$-map has enabled us to prove the somewhat complicated sign alternation patterns for the $\ell=2$ quasi-characters in rank 2.

\section{Comparison to mathematical literature}

\subsection{Kaneko-Zagier version of the $\Theta$-map}

The quasi-characters discussed above were motivated by earlier mathematical
studies of second-order MLDEs with the {\it non-zero Wronskian condition} (equivalent to $\ell=0$) \cite{KZ,KK,Kaneko:On}. However these equations were formulated for modular functions of specific non-zero weights rather than weight-0 VVMFs that are potential RCFT characters. The first example was the Kaneko-Zagier equation \cite{KZ} which is an $\ell=0$ MLDE for weighted VVMFs  that was studied in the context of super-singular elliptic curves. Subsequently \cite{Chandra:2018pjq} formulated  a ``dual Kaneko-Zagier equation'', which also acts on weighted modular functions and corresponds to the $\ell=2$ case.

The original Kaneko-Zagier equation is \cite{KZ}:
\begin{equation}
    \left(
        D^2-\frac{k(k+2)}{144}E_4
    \right)f_{(k)}=0,
    \label{KZeq_review}
\end{equation}
whose solutions have modular weight \(k\). Note the relation between the coefficient of $E_4$ and the weight of the solution. 
Under a change of variable:
\be
\chi^{(0)}=\eta^{-2k}f_{(k)}
\label{chvar}
\ee
Eq.~\eqref{KZeq_review} reduces to the MMS equation Eq.~\eqref{2ell0} with $c^{(0)}=2k$ and $h^{(0)}=\frac{k+1}{6}$. In this sense, KZ is just a reformulation of the MMS equation, though it was discovered independently by mathematicians.

Similarly the dual KZ equation is:
\begin{equation}
    \left(
        D^{2}
        +\frac{1}{3}\frac{E_6}{E_4}D
        -\frac{\widetilde k(\tk-2)}{144}E_4
    \right)\widetilde f_{(\tk)}=0.
    \label{dualKZeq_review}
\end{equation}
After setting
\(
\chi^{(2)}=\eta^{-2\tk}\widetilde f_{(\tk)}
\),
it reduces to the $\ell=2$ MLDE given in Eq.~\eqref{2ell2} with the following identifications: $c^{(2)}=2\tk$ and $h^{(2)}=\frac{\tk-1}{6}$.

Using the relation \eref{chvar} we can now work out the version of the $\Theta$-map that relates KZ and dual KZ solutions. We find that the map simplifies by losing its $\eta$-pre-factor, and relates 
weight-$k$ VVMFs $f_{(k)}$ solving Kaneko-Zagier equation to weight-$(k+2)$ VVMFs $\tf_{(k+2)}$ solving the dual Kaneko-Zagier equation in the following simple way:
\begin{equation}
    \tf_{(k+2)}
    \equiv D f_{(k)}
\end{equation}
Thus the $\Theta$-map has just reduced to the covariant derivative $D$! The proof is elementary and left as an exercise. Higher-order generalisations of Kaneko-Zagier-type differential equations, acting on solutions of specific non-zero weight, were studied in \cite{Kaneko:2017joa, nagatomo2022MLDE} and it should be straightforward to work out the analogous result for those cases. 

Note that while our differential operator $\Theta$ preserves the weight (which is zero) but introduces a multiplier system, the operator $D$ arising in the Kaneko-Zagier version changes the weight but preserve the multiplier system of the VVMF on which it acts.

\subsection{Bantay-Gannon differential operators}

We next compare the \(\Theta\)-map with the differential operators appearing in the Bantay-Gannon theory of vector-valued modular forms \cite{Bantay:2007zz, Bantay:2010uy, Gannon:2013jua}. For a fixed modular representation \(\rho\), Bantay and Gannon consider differential operators which act on the space of weakly holomorphic modular forms while preserving both the modular weight and the multiplier system. These operators are
\be\label{eq:BG_diff}
  \nabla_1
  =
  \frac{E_4E_6}{\Delta}D,
  \qquad
  \nabla_2
  =
  \frac{E_4^2}{\Delta}D^2,
  \qquad
  \nabla_3
  =
  \frac{E_6}{\Delta}D^3,
\ee
each of which has total weight zero. This is the crucial structural difference from the \(\Theta\)-map which, on the contrary, changes the multipler system.

It is easily verified that the Bantay-Gannon operators shift leading exponents by an integer, namely \cite{Bantay:2007zz}
\be
  \alpha\to \alpha-1,
\ee
and hence $c\to c+24$. 
Since the \(T\)-matrix only depends on the exponents modulo \(\mathbb Z\), this integer shift
does not change the multiplier system. This is exactly why for a fixed \(\rho\), these operators generate new weakly holomorphic VVMFs in the same representation.

On the other hand, for our $\Theta$-map, as we have seen above, the exponent shift is fractional:
\be
  \alpha\to \alpha-\frac16,
\ee
and hence $c\to c+4$. 
Consequently, the \(T\)-matrix is multiplied by \(v_\eta^{-4}\) and the \(S\)-matrix gets mapped to \(-S\).

Thus, the Bantay-Gannon differential operators as given in Eq.~\eqref{eq:BG_diff} are endomorphisms of a fixed space of VVMFs. They are designed to organise the space of VVMFs for a fixed modular representation \(\rho\). In contrast, the \(\Theta\)-map is not such an endomorphism, but rather a bridge between different multiplier systems:
\be
  \rho,\quad
  v_\eta^{-4}\otimes\rho,\quad
  v_\eta^{-8}\otimes\rho,\quad
  v_\eta^{-12}\otimes\rho, \quad \ldots ,
\ee
and therefore between different Wronskian index sectors.

\subsection{Bantay-Gannon duality}
\label{app:bg_duality}

We have seen in Section \ref{sec:r3}, that in the rank-3 case, the $\Theta$-map relates $\ell=0$ to $\ell=3$ MLDEs and their solutions. Let us compare this construction to another recent construction that also relates $\ell=0$ to $\ell=3$ solutions and vice-versa for the rank-3 case. It was introduced in \cite{Bantay:2007zz} and is referred to as {\it Bantay-Gannon duality} in \cite{Govindarajan:2025jlq}.

To set the stage, let us introduce some notation. Let $\mathbb{X}$ denote a set of admissible characters of rank $p$, and let $\Xi(q)$ denote the $p\times p$ ``fundamental matrix'' \cite{Bantay:2007zz, Gannon:2013jua} which has the following $q$-series: 
\be
\Xi(q)=q^\Lambda\left({\mathds 1}_p + \mathcal{Y}q+\mathcal{O}(q^2)\right),
\ee
This defines the $p\times p$ matrices $\Lambda$ and $\mathcal{Y}$ and fixes the normalisation. $\mathcal{Y}$ is a matrix with rational entries and $\Lambda$ is a diagonal matrix with entries: $\lambda_i\equiv\alpha_i\,\,(\text{mod }1)$ with the $\alpha_i$ being the exponents defined in Eq.~\eqref{qseries}.

Bantay-Gannon define a ``dual'' matrix by \cite{Govindarajan:2025jlq, Bantay:2007zz}:
\begin{equation}
\Xi^\vee(\tau)
\equiv
\frac{E_4(\tau)^2E_6(\tau)}{\Delta(\tau)^{7/6}}
\left(\Xi(\tau)^T\right)^{-1}.
\label{eq:BGdual}
\end{equation}
In the dual matrix, the above matrices change to:
\begin{equation}
\Lambda^\vee=-\frac76\,{\mathds 1}_p-\Lambda,
\qquad
\mathcal{Y}^\vee=4\cdot{\mathds 1}_p-\mathcal{Y}^T .
\label{eq:BGdata}
\end{equation}
In general, if \(\ell\) is the Wronskian index of $\mathbb{X}$, the $\mathbb{X}$ column of the dual has \cite{Govindarajan:2025jlq}:
\begin{equation}
\ell^\vee=(p-5)(p-1)+7-\ell ,
\label{eq:BGellgeneral}
\end{equation}
Now let us specialise to \(p=3\). Then this simplifies to:
\begin{equation}
\ell^\vee=3-\ell .
\label{eq:BGell3}
\end{equation}
Thus Bantay-Gannon duality Eq.~\eqref{eq:BGdual} gives a mapping between the bases: $(3,0) \leftrightarrow (3,3)$. Let us also note the transformation of the multiplier system. If:
\be
\Xi(\gamma\tau)=\rho(\gamma)\Xi(\tau),
\ee
then
\be
\left(\Xi(\gamma\tau)^T\right)^{-1}
=
\rho(\gamma)^{-T}\left(\Xi(\tau)^T\right)^{-1}.
\ee
The pre-factor in Eq.~\eqref{eq:BGdual} has weight zero and carries the multiplier \(v_\eta^{-4}\). Therefore
\begin{equation}
\rho^\vee(\gamma)=v_\eta^{-4}(\gamma)\otimes \rho(\gamma)^{-T}.
\label{eq:BGmult}
\end{equation}
Equivalently, at the level of the \(S\)-matrix,
\be
S^\vee=-(S^{-1})^T .
\ee

We can now compare this with the $\Theta$-map $(3,0)\leftrightarrow (3,3)$ where we instead had:
\be
\rho_\Theta(\gamma) = v_\eta^{-4}(\gamma)\otimes \rho(\gamma)
\ee
which implies $S\to -S$.

Thus the two constructions are not the same in general: $\rho^\vee\neq\rho_\Theta$. Also, for the $\Theta$-map construction we have a map at the level of $\ell=0$ and $\ell=3$ MLDEs whereas the Bantay-Gannon duality is a map between the $\ell=0$ and $\ell=3$ VVMFs.

It is useful to see why both constructions land in the same Wronskian index sector. Let the exponents of a $\ell=0$ solution be
\[
\alpha_0=-\frac{c}{24},\qquad
\alpha_1=-\frac{c}{24}+h_1,\qquad
\alpha_2=-\frac{c}{24}+h_2 .
\]
For the canonical basis one has \cite{Govindarajan:2025jlq}
\begin{equation}
\Lambda=\mathrm{diag}(\alpha_0,\alpha_1-1,\alpha_2-1).
\label{eq:rank3Lambda}
\end{equation}
Using Eq.~\eqref{eq:BGdata}, the three columns of \(\Xi^\vee\) have exponent
triples
\begin{equation}
\left(
-\frac76-\alpha_0,\,
\frac56-\alpha_1,\,
\frac56-\alpha_2
\right),
\label{eq:BGtriple1}
\end{equation}
\begin{equation}
\left(
-\frac16-\alpha_0,\,
-\frac16-\alpha_1,\,
\frac56-\alpha_2
\right),
\label{eq:BGtriple2}
\end{equation}
and
\begin{equation}
\left(
-\frac16-\alpha_0,\,
\frac56-\alpha_1,\,
-\frac16-\alpha_2
\right).
\label{eq:BGtriple3}
\end{equation}
Now the rank-3 \(\ell=0\) valence formula gives:
\be
\alpha_0+\alpha_1+\alpha_2=\frac12,
\ee
so each of the triples in Eqs.~\eqref{eq:BGtriple1}--\eqref{eq:BGtriple3} has vanishing sum. Hence each column has \(\ell=3\). On the other hand, the \(\Theta\)-map acts by the uniform shift:
\begin{equation}
\alpha_i\to \alpha_i-\frac16 ,
\label{eq:Theta_uniform_shift}
\end{equation}
which also gives
\be
\sum_{i=0}^2\left(\alpha_i-\frac16\right)=0.
\ee
Thus both constructions send rank-three $\ell=0$ data to the $\ell=3$ Wronskian index sector, but the associated multiplier systems are different.

Let us also note that Bantay-Gannon duality maps $(p,0)$ to a rank $p$ VVMF with $\ell$ growing quadratically in $p$, as seen from Eq.~\eqref{eq:BGellgeneral}. On the contrary, the $\Theta$-map, as has been noted before, maps $(p,0)$ to $(p,p)$. For low values of $p$ there are a couple of coincidences: at $p=3,4$ both approaches send $(p,0)\to(p,p)$. However at rank 5, Bantay-Gannon duality maps $(5,0)$ to $(5,7)$ while the $\Theta$-map takes $(5,0)$ to $(5,5)$ and the difference between the two grows rapidly with $p$. Hence, the programme of classifying quasi-characters at arbitrary Wronskian index in terms of those with $\ell=0$ seems more effective via the use of the $\Theta$-map.

\section{Discussion}

We have proposed a novel differential map which we call the $\Theta$-map that relates infinitely many MLDEs, as well as their quasi-character solutions, to the simplest MLDEs of each rank namely those with vanishing Wronskian index ($\ell=0$). This amounts to a major simplification in the classification of general families of quasi-characters, among which all admissible characters necessarily lie. In the case of rank 2 and 3, all MLDEs get related to the simplest ones: the MMS equation \cite{Mathur:1988na} for rank-2 and the analogous $\ell=0$ equation first studied in \cite{Mathur:1988gt} for rank-3.

A particularly tractable next step is the complete classification of $(3,0)$ quasi-characters. Such a result would provide for rank 3 a level of control comparable to that already available in rank 2. Since the allowed Wronskian indices in rank 3 are multiples of three, repeated applications of the generalised $\Theta$-operators constructed here generate all higher-$\ell$ sectors from $(3,0)$ data. Thus, within the class of solutions considered in this work (non-logarithmic and having an irreducible modular representation), the classification problem across the rank-3 Wronskian tower is reduced to the classification of this vanishing Wronskian index sector.

The sign and growth results for the rank-2 $\ell=0$ families established in \cite{Das:2025gto}, together with the $\Theta$-map developed here, has enabled us to prove sign and growth results for the rank-2 $\ell=2$ families which had so far only been conjectured. 
These analyses are already beginning to find further applications. In particular, the $q$-series that were studied in \cite{Alexandrov:2023ltz, Alexandrov:2023zjb} exhibit sign alternation that was conjectured to terminate after finitely many terms. The quantitative control over the signs and growth of quasi-character coefficients developed in the present work and \cite{Das:2025gto} now provides the ingredients needed to address this sign behaviour, and ongoing work \cite{DasMukhi:unpub} indicates that it should be possible to establish the above conjectures in this way. The significance of proving this sign stabilisation is not merely arithmetic. The coefficients of the (mock) modular generating series in \cite{Alexandrov:2023ltz,Alexandrov:2023zjb} are indexed degeneracies of D4--D2--D0 BPS states and hence encode microscopic black-hole states, while their signs reflect cancellations intrinsic to the supersymmetric index. More recently, \cite{Alexandrov:2026rra} observed that changes in the sign behaviour of Gopakumar-Vafa (GV), Pandharipande-Thomas (PT) and Donaldson-Thomas (DT) invariants, as well as of the five-dimensional BPS index, accompany transitions between different dominant macroscopic configurations. In particular, for PT invariants the alternating pattern terminates at the first kink across which the dominant contribution changes from multi-centred bound states to single-centred black holes, while the corresponding five-dimensional kink separates the BMPV-black-hole \cite{Breckenridge:1996is} and black-ring regimes. Proving that the alternation terminates, and determining precisely where this occurs, therefore gives rigorous control over a microscopic diagnostic of these transitions and of the cancellations preceding the macroscopic black-hole regime.

For ranks $p\geq4$, the natural extension of this programme is to classify the finite set of Wronskian sectors with $0\leq\ell<p$, since the $\Theta$-map then organises each such sector into a tower with $\ell\to\ell+p$. It would also be important to determine systematically when the resulting quasi-characters, or suitable linear combinations of them, are admissible, since the map does not in general preserve positivity. Extending the sign and growth estimates to higher Wronskian index and higher rank, together with an analysis of the exceptional and logarithmic cases of the inverse map, should provide useful criteria for isolating new candidate RCFT characters within these towers. A complementary direction is to seek a conformal-field-theoretic or representation-theoretic interpretation of the multiplier-system induced by the $\Theta$-map and to clarify further its relation to other differential operators \cite{Bantay:2007zz, Govindarajan:2025jlq, nagatomo2022MLDE}.

\section*{Acknowledgements}

The work of AD is supported by the STFC Consolidated Grant ST/T000600/1 ``Particle Theory at the Higgs Centre''. SM is supported by the Raja Ramanna Chair of the Department of Atomic Energy, Government of India. We are grateful to Suresh Govindarajan, Chethan Gowdigere and Jagannath Santara for useful discussions regarding the rank three MLDE. AD also thanks Arnab Chakraborty and Jishu Das for insightful discussions on modular forms.

\appendix

\section{Useful modular identities}

Let $E_4,E_6$ be the Eisenstein series of the corresponding modular weight, $\eta$ be the Dedekind eta-function, and $j$ be the Klein $j$-invariant with expansion:
\be
j(q)=q^{-1}+744+\ldots
\ee
The following identities among these objects are standard and were used in this paper:
\be\label{j_ids}
\begin{split}
&q\frac{d}{dq} E_2 = \frac{1}{12}(E_2^2-E_4), \qquad DE_4 =-\frac13 E_6, \qquad DE_6 =-\half E_4^2 \qquad \hbox{(Ramanujan identities)}\\
& D\eta =0, \quad\quad\qquad\quad\qquad\quad \, \, \, \, j^\frac13 =\eta^{-8}E_4,  \quad\qquad \, \, \, \, (j-1728)^{\frac12} = \eta^{-12}E_6\\
& Dj =-\frac{E_6}{E_4}j, \quad\qquad\quad\qquad  D^2j = \left(\half E_4+\frac{2}{3}\frac{E_6^2}{E_4^2}\right)j
\end{split}
\ee
 Next we recall some identities involving the Jacobi theta functions:
\begin{equation}
    \theta_2(\tau)^2
    =
    2\theta_2(2\tau)\theta_3(2\tau),
    \qquad
    \theta_3(\tau)^2
    =
    \theta_3(2\tau)^2+\theta_2(2\tau)^2,
    \qquad
    \theta_4(\tau)^2
    =
    \theta_3(2\tau)^2-\theta_2(2\tau)^2,
    \label{thetaPn_duplication}
\end{equation}
and
\begin{equation}
    2\eta(\tau)^3
    =
    \theta_2(\tau)\theta_3(\tau)\theta_4(\tau), \qquad \theta_3^4 = \theta_2^4 + \theta_4^4.
    \label{thetaPn_jacobi}
\end{equation}
We will also need identities involving derivatives of Jacobi theta functions:
\be\label{eq:theta_derive}
    \frac{1}{2\pi i}\partial_\tau\log\frac{\theta_2(\tau)}{\theta_3(\tau)} = \frac{1}{24}(\theta_3^4 - \theta_2^4 + 2\theta_4^4)
\ee

\section{Inverses for the generic $\Theta$-map}\label{sec:gen_inv}

In this Appendix we explain how to find the inverses for the generic $\Theta$-maps given in \eref{genThetaop}. To begin, let us consider the $\Theta$-map relation rewritten as:
\begin{equation}
    D\chi^{(\ell)}
    =
    \eta^4\chi^{(\ell+p)} .
    \label{forward_relation_for_inverse}
\end{equation}
Since \(D\eta=0\), repeated differentiation gives
\begin{equation}
    D^m\chi^{(\ell)}
    =
    \eta^4D^{m-1}\chi^{(\ell+p)},
    \qquad m\geq 1.
    \label{Dm_inverse_relation}
\end{equation}
Now the MLDE for Wronskian index $\ell$ has the form:
\begin{equation}
    \left(
        D^p
        + \phi_2D^{p-1}
        + \phi_4D^{p-2}
        +\cdots
        + \phi_{2p-2}D
        + \phi_{2p}
    \right)\chi^{(\ell)}=0.
    \label{general_source_mlde_for_inverse}
\end{equation}
Using Eq.~\eqref{Dm_inverse_relation}, all terms except the last one can be rewritten in terms of \(\chi^{(\ell+p)}\):
\begin{equation}
    \phi_{2p}\chi^{(\ell)}
    =
    -\eta^4
    \left(
        D^{p-1}
        + \phi_2D^{p-2}
        + \phi_4D^{p-3}
        +\cdots
        + \phi_{2p-2}
    \right)\chi^{(\ell+p)} .
\end{equation}
Therefore, whenever \(\phi_{2p}\neq 0\), the inverse map is
\begin{equation}
    \chi^{(\ell)}
    =
    -\frac{\eta^4}{\phi_{2p}}
    \left(
        D^{p-1}
        + \phi_2D^{p-2}
        + \phi_4D^{p-3}
        +\cdots
        + \phi_{2p-2}
    \right)\chi^{(\ell+p)} .
    \label{general_theta_inverse_rule}
\end{equation}
This is the inverse for the \(\Theta\)-map acting on a solution of generic $\ell$. The idea is to remove the zero-order term of the source MLDE, act with the remaining differential operator on the image, and divide by the zero-order coefficient.

As a check, let us look at the rank-3 case. We recover the inverse for the $\ell=0$ to $\ell=3$ map, given in \eref{rank3_inverse_AB}, by taking:
\be
p=3, \quad \phi_2 = 0, \quad \phi_4 = \mu_1\,E_4, \quad \phi_6 = \mu_2\,E_6.
\ee

Let us find the inverse of the operator given in Eq.~\eqref{genThetaop} by doing a simple linear algebra exercise below. To begin, let us denote it by $\mathcal{T}_n$:
\be
  \mathcal T_n
  =
  \sum_{2p+3q+r=n}
  X_{pqr}\,
  j^{p/3}(j-1728)^{q/2}\Theta^r
\ee
shifts every exponent by \(-n/6\), and hence sends $(p,\ell)\to (p,\ell+pn)$. Now consider the generic rank-\(p\) arbitrary $\ell$ MLDE given in Eq.~\eqref{general_source_mlde_for_inverse}. Let us re-write it as:
\be
  D^p\chi^{(\ell)}
  =
  -\sum_{r=0}^{p-1} A_r D^r\chi^{(\ell)},
  \qquad
  A_r\equiv \phi_{2(p-r)}.
\ee
For simplicity let us introduce the following column vector
\be
  \mathbf X
  =
  \begin{pmatrix}
    \chi^{(\ell)}\\
    D\chi^{(\ell)}\\
    \vdots\\
    D^{p-1}\chi^{(\ell)}
  \end{pmatrix}.
\ee
Then Eq.~\eqref{general_source_mlde_for_inverse} becomes:
\be\label{eq:DCX}
  D\mathbf X=\mathsf C\,\mathbf X,
\ee
where \(\mathsf C\) is given as:
\be
  \mathsf C\equiv
  \begin{pmatrix}
    0 & 1 & 0 & \cdots & 0\\
    0 & 0 & 1 & \cdots & 0\\
    \vdots & \vdots & \vdots & \ddots & \vdots\\
    0 & 0 & 0 & \cdots & 1\\
    -A_0 & -A_1 & -A_2 & \cdots & -A_{p-1}
  \end{pmatrix}.
\ee

For every non-negative integer \(r\), define a row vector \(\mathbf R_r\) by
\be\label{eq:R_defn}
  D^r\chi=\mathbf R_r\,\mathbf X.
\ee
For \(0\leq r\leq p-1\), \(\mathbf R_r\) is the standard basis row vector with a single non-zero entry in position \(r\).  For all \(r\geq 0\), then we obtain the following recursion relation (using Eq.~\eqref{eq:DCX} and derivative of Eq.~\eqref{eq:R_defn} ):
\be
  \mathbf R_{r+1}
  =
  D\mathbf R_r+\mathbf R_r\,\mathsf C.
\ee
Now using modular identities given in Eq.~\eqref{j_ids} we can rewrite the action of \(\mathcal T_n\) on the solution space as
\be
  \psi^{(\ell+pn)}\equiv \mathcal T_n\chi^{(\ell)}
  =
  \eta^{-4n}\,\mathbf F_n\,\mathbf X,
\ee
where
\be
  \mathbf F_n
  =
  \sum_{2p+3q+r=n}
  X_{pqr}\,E_4^p E_6^q\,\mathbf R_r.
\ee
The \(i\)-th component of \(\mathbf F_n\) has modular weight \(2n-2i\), so that
\(\mathbf F_n\mathbf X\) has total weight \(2n\), as required. We now differentiate: $\eta^{4n}\psi^{(\ell+pn)}=\mathbf F_n\mathbf X$. Define row vectors \(\mathbf F_n^{[m]}\) recursively by
\be
  \mathbf F_n^{[0]}\equiv\mathbf F_n,
  \qquad
  \mathbf F_n^{[m+1]}
  \equiv
  D\mathbf F_n^{[m]}+\mathbf F_n^{[m]}\mathsf C.
\ee
Then we get,
\be
  D^m(\eta^{4n}\psi^{(\ell+pn)})
  = \eta^{4n}D^m\psi^{(\ell+pn)} =
  \mathbf F_n^{[m]}\mathbf X,
  \qquad
  m=0,\ldots,p-1,
\ee
which can be written in matrix-form as:
\be\label{eq:psi_inv}
  \eta^{4n}
  \begin{pmatrix}
    \psi^{(\ell+pn)}\\
    D\psi^{(\ell+pn)}\\
    \vdots\\
    D^{p-1}\psi^{(\ell+pn)}
  \end{pmatrix}
  =
  \mathsf M_n
  \begin{pmatrix}
    \chi^{(\ell)}\\
    D\chi^{(\ell)}\\
    \vdots\\
    D^{p-1}\chi^{(\ell)}
  \end{pmatrix},
\ee
where
\be
  \mathsf M_n
  \equiv
  \begin{pmatrix}
    \mathbf F_n^{[0]}\\
    \mathbf F_n^{[1]}\\
    \vdots\\
    \mathbf F_n^{[p-1]}
  \end{pmatrix}.
\ee
Therefore, whenever $\det\mathsf M_n\neq 0$, Eq.~\eqref{eq:psi_inv} can be inverted. In particular, the inverse for the original character $\chi^{(\ell)}$ can be obtained from the first component above:
\be
  \chi^{(\ell)}
  =
  \eta^{4n}\,
  e_0^{\,T}\mathsf M_n^{-1}
  \begin{pmatrix}
    \psi^{(\ell+pn)}\\
    D\psi^{(\ell+pn)}\\
    \vdots\\
    D^{p-1}\psi^{(\ell+pn)}
  \end{pmatrix},
\ee
where
\be
  e_0=(1,0,\ldots,0)^T.
\ee

\section{$\Theta$-map relations for the elliptic-function
representation of quasi-characters}\label{app:theta_Pn}

In the papers on the rank-2 Kaneko-Zagier equation, certain recursive polynomials were found that describe the solutions of these equations \cite{KK, Kaneko:On, Kaneko:2013uga}. Correspondingly one finds recursive polynomials that describe solutions of the corresponding weight-0 VVMFs with $\ell=0$, these are described in \cite{Chandra:2018pjq} together with their $\ell=2$ generalisations. Here we relate these polynomials using the $\Theta$-map. For simplicity, we work with the $\mA_{1}$, dual $\mA_1$ and $\mE_{7}$, dual $\mE_7$ families of quasi-characters. The remaining families can be worked out similarly\,\footnote{There is a difference in the way families of quasi-characters are labelled in \cite{Das:2025gto} and \cite{Chandra:2018pjq}. Here, we are following the conventions of \cite{Das:2025gto} and thus $\mA_1$ and $\mE_7$ fall under different families while in \cite{Chandra:2018pjq} they belonged to the same family of quasi-characters.}.

Since the notation $\chi_i^{(\ell)}$ is reserved for the quasi-characters themselves, we denote the two characters of the $\mA_{1,1}$ theory by \cite{Chandra:2018pjq}:
\begin{equation}
    \psi_0(\tau)
    \equiv
    \frac{\theta_3(2\tau)}{\eta(\tau)},
    \qquad
    \psi_1(\tau)
    \equiv
    \frac{\theta_2(2\tau)}{\eta(\tau)},
    \qquad
    t(\tau)
    \equiv
    \left(\frac{\psi_1}{\psi_0}\right)^4
    ,
    \label{thetaPn_basic_defs}
\end{equation}
where $\theta_2$, $\theta_3$ are the Jacobi theta-functions\,\footnote{Notice that $t(\tau)=\lambda(2\tau)$ where $\lambda$ is the modular lambda function.}. It is useful to introduce the integers to relate the notation of \cite{Chandra:2018pjq} with that of \cite{Das:2025gto} for the quasi-characters labels:
\begin{equation}
    \delta_j
    \equiv
    \frac{j-1}{6},
    \qquad
    \nu_{M,j}
    \equiv
    4M+\delta_j,
    \qquad
    j\in\{1,7\},
    \qquad
    M\in\mathbb{Z}_{\geq0}.
    \label{thetaPn_index_translation}
\end{equation}
Thus
\begin{equation}
    \delta_1=0,
    \qquad
    \delta_7=1,
    \qquad
    24M+j=6\nu_{M,j}+1,
    \label{thetaPn_index_relations}
\end{equation}
where for notational simplicity we have identified: $(M,j)\equiv (M^{(0)},j^{(0)})$ for this appendix only. The cases \(j=1\) and \(j=7\) above correspond respectively to the \(\mA_1\) and \(\mE_7\) families of \(\ell=0\) quasi-characters as described earlier in Section \ref{sec:quasi2}. In terms of the \(r\)-indexed polynomials of \cite{Chandra:2018pjq}, we have the following dictionary:
\be
 P^{(0)}_{M,j}(t)
 \equiv P^{(0)}_{\nu_{M,j}}(t),
 \qquad
 P^{(2)}_{M,j}(t)
 \equiv P^{(2)}_{\nu_{M,j}+1}(t).
\label{thetaPn_polynomial_definition}
\ee
The \((M,j)\)-indexed polynomials below are simply the subsequences of the
original \(r\)-indexed polynomials in \cite{Chandra:2018pjq} adapted to the quasi-character labels of \cite{Das:2025gto}. In Eq.~\eqref{thetaPn_polynomial_definition} $r\equiv \nu_{M,j}$ and $r\equiv \nu_{M,j}+1$ for the $\ell=0, 2$ families of quasi-characters respectively. Now Eq.~\eqref{thetaPn_index_translation} gives
\be
 P^{(0)}_{M,1}= P^{(0)}_{4M},
 \qquad
 P^{(0)}_{M,7}= P^{(0)}_{4M+1},
 \qquad
 P^{(2)}_{M,1}= P^{(2)}_{4M+1},
 \qquad
 P^{(2)}_{M,7}= P^{(2)}_{4M+2}.
\ee
The initial polynomials follow by evaluating the above cases at
\(M=0,1\), while the recursion relations below are obtained by restricting the original \(r\)-indexed sequences of \cite{Chandra:2018pjq} to these families and applying the original recursions twice so that they close for each fixed value of \(j\).

Let us introduce the polynomial sequence
\(P^{(0)}_{M,j}(t)\) through:
\be
\begin{aligned}
 P^{(0)}_{0,1}(t)
   &=1,
 \qquad\qquad P^{(0)}_{1,1}(t)
   =1-20t+630t^2+2380t^3+1105t^4,
 \\[2mm]
 P^{(0)}_{0,7}(t)&=1+7t
 \qquad
 P^{(0)}_{1,7}(t)
   =1-\frac{155}{7}t+\frac{2790}{7}t^2
     +\frac{64170}{7}t^3+\frac{121923}{7}t^4
     +\frac{40641}{7}t^5 .
\end{aligned}
\label{thetaPn_source_poly_initial}
\ee
For \(M\geq 2\), these polynomials are defined recursively by
\be
\begin{aligned}
 P^{(0)}_{M,j}(t)
 ={}&
 \left[
 (t+1)^2(t^2-34t+1)^2
 +\frac{
 6\bigl(36\nu_{M,j}^{\,2}-252\nu_{M,j}+317\bigr)
 }{
 (2\nu_{M,j}-11)(2\nu_{M,j}-3)
 }
 t(t-1)^4
 \right]
 P^{(0)}_{M-1,j}(t)
 \\[1mm]
 &\,
 -\frac{
 9(6\nu_{M,j}-43)(6\nu_{M,j}-35)
  (6\nu_{M,j}-31)(6\nu_{M,j}-23)
 }{
 (2\nu_{M,j}-15)(2\nu_{M,j}-11)^2
 (2\nu_{M,j}-7)
 }
 t^2(t-1)^8
 P^{(0)}_{M-2,j}(t).
\end{aligned}
\label{thetaPn_source_poly_recursion}
\ee
Similarly, for the \(\ell=2\) dual \(\mA_1\) and dual \(\mE_7\) families, we have
the polynomial sequence \(P^{(2)}_{M,j}(t)\):
\be
\begin{aligned}
 P^{(2)}_{0,1}(t)
   &=1-5t,
 \qquad\qquad\qquad
 P^{(2)}_{1,1}(t)
   =1-\frac{29}{5}t-\frac{2494}{5}t^2
     -\frac{32074}{5}t^3-\frac{40919}{5}t^4
     -\frac{6409}{5}t^5,
 \\[2mm]
 P^{(2)}_{0,7}(t)
   &=1-22t-11t^2, \qquad
 P^{(2)}_{1,7}(t)
   =1-10t-125t^2-13500t^3-61065t^4
     -49818t^5-6555t^6 .
\end{aligned}
\label{thetaPn_dual_poly_initial}
\ee
For \(M\geq 2\), these polynomials are defined recursively by
\be
\begin{aligned}
 P^{(2)}_{M,j}(t)
 ={}&
 \left[
 (t+1)^2(t^2-34t+1)^2
 +\frac{
 6\bigl(36\nu_{M,j}^{\,2}-252\nu_{M,j}+269\bigr)
 }{
 (2\nu_{M,j}-11)(2\nu_{M,j}-3)
 }
 t(t-1)^4
 \right]
 P^{(2)}_{M-1,j}(t)
 \\[1mm]
 &\,
 -\frac{
 9(6\nu_{M,j}-47)(6\nu_{M,j}-35)
  (6\nu_{M,j}-31)(6\nu_{M,j}-19)
 }{
 (2\nu_{M,j}-15)(2\nu_{M,j}-11)^2
 (2\nu_{M,j}-7)
 }
 t^2(t-1)^8
 P^{(2)}_{M-2,j}(t).
\end{aligned}
\label{thetaPn_dual_poly_recursion}
\ee

In \cite{Chandra:2018pjq} it is shown that the $\ell=0$ quasi-characters can be written in terms of the above objects as:
\begin{equation}
    \chi^{(0)}_{0}(\tau)
    =
    \psi_0(\tau)^{24M+j}
    P^{(0)}_{M,j}(t),
    \qquad
    \chi^{(0)}_{1}(\tau)
    =
    \psi_1(\tau)^{24M+j}
    P^{(0)}_{M,j}(t^{-1}),
    \label{thetaPn_A1_series}
\end{equation}
with
\begin{equation}
    c^{(0)}_{M,j}
    =
    24M+j,
    \qquad
    h^{(0)}_{M,j}
    =
    2M+\frac{j+2}{12}.
    \label{thetaPn_A1_ch}
\end{equation}
The corresponding dual $\ell=2$ quasi-characters become
\begin{equation}
    \chi^{(2)}_{0}(\tau)
    =
    \psi_0(\tau)^{24M+j+4}
    P^{(2)}_{M,j}(t),
    \qquad
    \chi^{(2)}_{1}(\tau)
    =
    \psi_1(\tau)^{24M+j+4}
    P^{(2)}_{M,j}(t^{-1}),
    \label{thetaPn_dual_A1_series}
\end{equation}
with
\begin{equation}
    c^{(2)}_{M,j}
    =
    24M+j+4,
    \qquad
    h^{(2)}_{M,j}
    =
    2M+\frac{j+2}{12}.
    \label{thetaPn_dual_A1_ch}
\end{equation}
It follows from the results of Section~\ref{sec:rk2} and
Eqs.~\eqref{thetaPn_A1_series}--\eqref{thetaPn_dual_A1_series} that, under the \(\Theta\)-map \footnote{Here the labels \((M,j)\) on the dual quasi-characters are inherited from the \(\ell=0\) family, and the two components have not yet been exchanged. After the exchange used in Section~\ref{sec:rk2}, the corresponding standard dual-family labels are as given in Eq.~\eqref{eq:M2_M0}.}:
\begin{equation}
    \Theta\chi^{(0)}_{0}
    =
    -\frac{24M+j}{24}\,
    \chi^{(2)}_{0},
    \qquad
    \Theta\chi^{(0)}_{1}
    =
    -\frac{24M+j}{24}\,
    \chi^{(2)}_{1},
    \label{thetaPn_main_map}
\end{equation}
where, above, the $\ell=0$ quasi-character labelled by $(M,j)$ is mapped
to the corresponding dual $\ell=2$ quasi-character carrying the
same labels. Using the above result, we will find the relations between these recursively defined polynomial sequences.

From the identities involving Jacobi-theta functions, we see that, squaring the first equation in Eq.~\eqref{thetaPn_jacobi} and using Eq.~\eqref{thetaPn_duplication}, one obtains
\begin{equation}
    \psi_0^6
    =
    2t^{-1/4}(1-t)^{-1},
    \qquad
    \psi_1^6
    =
    2t^{5/4}(1-t)^{-1}.
    \label{chi_01_6}
\end{equation}
Next let us compute the action of $\Theta$ in the $t$-variable. Using Eq.~\eqref{eq:theta_derive}, one has
\begin{equation}
    Dt
    =
    \theta_3(2\tau)^4t(1-t).
    \label{thetaPn_Dt}
\end{equation}
Now let us multiply above by $\eta^{-4}$. Using the first equation in Eq.~\eqref{thetaPn_basic_defs}, we get
\begin{equation}
    \Theta t
    =
    \psi_0^4t(1-t), \quad \text{and thus, on functions of }t, \quad \Theta
    =
    \psi_0^4t(1-t)\partial_t.
    \label{thetaPn_Theta_t}
\end{equation}
From Eq.~\eqref{chi_01_6}, we obtain
\begin{equation}
    \partial_t\psi_0
    =
    -\frac{1-5t}{24t(1-t)}\psi_0,
    \qquad
    \partial_t\psi_1
    =
    \frac{5-t}{24t(1-t)}\psi_1.
    \label{thetaPn_dchi}
\end{equation}
Applying Eq.~\eqref{thetaPn_Theta_t} to the identity character in
Eq.~\eqref{thetaPn_A1_series}, we find
\be
\begin{aligned}
    \Theta\chi^{(0)}_{0}
    &=
    \psi_0^4t(1-t)\partial_t
    \left(
        \psi_0^{24M+j}P^{(0)}_{M,j}(t)
    \right)
    \\
    &=
    \psi_0^{24M+j+4}
    \left[
        t(1-t)
        \bigl(P^{(0)}_{M,j}\bigr)'(t)
        -
        \frac{24M+j}{24}
        (1-5t)P^{(0)}_{M,j}(t)
    \right], \\
    &= \psi_0(\tau)^{24M+j+4}\left[-\frac{24M+j}{24}
    P^{(2)}_{M,j}(t)\right]
    \label{thetaPn_Theta_Xn}
\end{aligned}
\ee
where the prime denotes differentiation with respect to the argument $t$ and to get to the last equality above we have used Eqs.~\eqref{thetaPn_dual_A1_series}--\eqref{thetaPn_main_map}. Thus, the polynomial identity induced by the $\Theta$-map is
\begin{equation}
    P^{(2)}_{M,j}(t)
    =
    (1-5t)P^{(0)}_{M,j}(t)
    -
    \frac{24}{24M+j}\,
    t(1-t)
    \bigl(P^{(0)}_{M,j}\bigr)'(t).
    \label{thetaPn_poly_identity}
\end{equation}
For the non-identity character, a similar computation yields the same expression as above with $t\leftrightarrow t^{-1}$. 

Thus the $\Theta$-map directly tells us how to obtain the polynomials $ P^{(2)}_{M,j}(t)$ for the dual family if we know  $P^{(0)}_{M,j}(t)$ for the original family.

\section{No-logarithm condition for the \((3,3)\) MLDE}\label{app:loc_reg}

Here we give the details leading to the no-logarithm condition used in Eq.~\eqref{rank3_nolog_condition} for the rank-3 MLDE. This condition is obtained by demanding local regularity of the solutions at $\tau=i$ and thus helps in reducing the number of parameters in the MLDE in Eq.~\eqref{rank3_33_raw} from three to two. Previously, in \cite{Gowdigere:2023xnm, Govindarajan:2025jlq} the $\ell=3$ MLDE had been studied and its admissible solutions had also been found. However, they had to work with the full three-parameter MLDE and it was noted in \cite{Gowdigere:2023xnm} (see Eq.~(58)) that all these solutions lie in a two-dimensional plane in the space spanned by these three parameters. We show below that this two-dimensional plane is obtained by demanding local regularity of the solutions around $\tau=i$.

We start directly from the general $\ell=3$ MLDE in Eq.~\eqref{rank3_33_raw},
\begin{equation}
\left[
D^3+\frac{1}{2}\frac{E_4^2}{E_6}D^2+\nu_1E_4D
+\nu_2E_6+\nu_3\frac{\Delta}{E_6}
\right]\chi_i^{(3)}=0 .
\label{locreg_33_raw_again}
\end{equation}
The elliptic point \(\tau= i\) corresponds to \(j=1728\). Using the identities
in Eq.~\eqref{j_ids}, together with \(E_4^3-E_6^2=1728\Delta\),
Eq.~\eqref{locreg_33_raw_again} becomes, in the \(j\)-coordinate,
\begin{equation}
\begin{split}
\Bigg[
\partial_j^3
+\left(\frac{2}{j}+\frac{1}{j-1728}\right)\partial_j^2
&+\left(
\frac{2}{9j^2}
+\frac{1+\nu_1}{j(j-1728)}
-\frac{1}{4(j-1728)^2}
\right)\partial_j  -\frac{\nu_2(j-1728)+\nu_3}{j^2(j-1728)^2}
\Bigg]\chi(j)=0 ,
\end{split}
\label{locreg_33_jcoord}
\end{equation}
where we have identified \(\chi(j)\equiv\chi_i^{(3)}(\tau(j))\).

We now study the Frobenius expansion about \(j=1728\). Let
\begin{equation}
\chi(j)=\sum_{n=0}^{\infty}f_n (j-1728)^{n+\lambda},
\qquad f_0\neq0 .
\label{locreg_frob_ansatz}
\end{equation}
The coefficient of \((j-1728)^{\lambda-3}\) gives the indicial equation for \eref{locreg_33_jcoord},
\begin{equation}
\lambda(\lambda-1)(\lambda-2)
+\lambda(\lambda-1)-\frac{1}{4}\lambda=0 ,
\label{locreg_indicial_eq}
\end{equation}
whose roots are: $\lambda = 0, \frac{1}{2}, \frac{3}{2}$. Thus the last two exponents differ by an integer. This is the reason why in the Frobenius expansion, the solution with the smaller exponent \(\lambda=\frac12\) can develop a logarithmic partner. The absence of this logarithmic solution would impose a compatibility condition at the next order. To see this explicitly, let us expand the singular coefficients of
Eq.~\eqref{locreg_33_jcoord} near \(j=1728\):
\begin{equation}
\begin{split}
&\frac{2}{j}+\frac{1}{j-1728}
=
\frac{1}{j-1728}+\frac{2}{1728}+\cO(j-1728), \\
&\frac{2}{9j^2}
+\frac{1+\nu_1}{j(j-1728)}
-\frac{1}{4(j-1728)^2}
=
-\frac{1}{4(j-1728)^2}
+\frac{1+\nu_1}{1728(j-1728)}
+\cO(1), \\
&-\frac{\nu_2(j-1728)+\nu_3}{j^2(j-1728)^2}
=
-\frac{\nu_3}{1728^2(j-1728)^2}
+\cO\!\left((j-1728)^{-1}\right).
\end{split} \label{locreg_expand_A}
\end{equation}
At the next order, namely at order \((j-1728)^{\lambda-2}\), the recursion
relation is
\begin{equation}
\begin{split}
&\left[
(\lambda+1)\lambda(\lambda-1)
+(\lambda+1)\lambda
-\frac14(\lambda+1)
\right]f_1 
+\left[
\frac{2}{1728}\lambda(\lambda-1)
+\frac{1+\nu_1}{1728}\lambda
-\frac{\nu_3}{1728^2}
\right]f_0=0 .
\end{split}
\label{locreg_next_order}
\end{equation}
The coefficient of \(f_1\) is the indicial polynomial
\eqref{locreg_indicial_eq} evaluated at \(\lambda+1\). For
\(\lambda=\frac12\), this means evaluating the indicial polynomial at
\(\lambda+1=\frac32\), which is again a root. Hence the coefficient of
\(f_1\) vanishes. Thus the recursion does not determine \(f_1\). Instead, the remaining \(f_0\)-term must vanish. Substituting \(\lambda=\frac12\) in Eq.~\eqref{locreg_next_order}, we get
\begin{equation}
\left(
\frac{\nu_1}{3456}
-\frac{\nu_3}{1728^2}
\right)f_0=0 .
\label{locreg_compatibility}
\end{equation}
Since \(f_0\neq0\), local regularity at \(j=1728\) requires
\begin{equation}
\nu_3=864\,\nu_1 .
\label{locreg_nolog_final}
\end{equation}
This is precisely the no-logarithm condition quoted in Eq.~\eqref{rank3_nolog_condition}.

\section{Some relevant mathematical estimates}\label{app:estimates_log}

In this Appendix, we record some mathematical estimates and results which we will need to prove the sign patterns for the $\ell=2$ quasi-characters as discussed in Section \ref{sec:l2-signs-theta}.

Let us start by writing the $q$-series of the Dedekind $\eta$-function as:
\begin{equation}
        \eta^{-4}
        =
        q^{-1/6}E(q),
        \qquad
        E(q)
        \equiv
        \prod_{s\ge1}(1-q^s)^{-4}
        =
        \sum_{s\ge0}E_sq^s,
\label{eq:E-kernel-defa}
\end{equation}
with $E_0=1$ and $E_s>0\,\,(\text{for }s\ge1)$. Now let us consider the $q$-series of an arbitrary weight zero modular function and then act on it with the Ramanujan-Serre derivative to get,
\be
        F(q)=q^\alpha\sum_{n\ge0}F_nq^n, \qquad D\,F(q)=q^\alpha\sum_{n\ge0}(\alpha+n)F_nq^n.
\ee
Now multiplying the second equation above by \(\eta^{-4}\), we get
\begin{equation}
\eta^{-4}D\,F(q)
=
q^{\alpha-\frac16}
\sum_{n\ge0}
\left(
\sum_{s=0}^{n}
E_s(\alpha+n-s)F_{n-s}
\right)q^n = q^{\alpha-\frac16}
\sum_{n\ge0}
\left(
\sum_{s=0}^{n}
E_s\,d_{n-s}
\right)q^n.
\label{eq:convolution-formula}
\end{equation}
with $d_l\equiv(\alpha+l)F_l$.

\subsection{Results on coloured partitions}

The coefficients \(E_s\) have a standard combinatorial interpretation -- it is the \(4\)-coloured partitions of \(s\) and is usually denoted by: \(p_4(s)\) (see \cite{BringmannKanePahariRolenStrictLC} for more details). Now let us use the {\it log-concavity theorem} for \(k\)-coloured partitions \cite{BringmannKanePahariRolenStrictLC}. In the case needed
here, \(k=4\), it says that \(p_4(s)\) is log-concave:
\[
        p_4(s)^2\ge p_4(s-1)p_4(s+1)
        \qquad(s\ge1).
\]
Equivalently,
\[
        E_s^2\ge E_{s-1}E_{s+1}
        \qquad(s\ge1).
\]
Now let us show that:
\be\label{eq:ker_esa}
        E_s\le 4\left(\frac72\right)^{s-1}
        \qquad(s\ge1).
\ee
To see this, let us define the following non-increasing ratio: $R_s\equiv\frac{E_s}{E_{s-1}}$. Now since $E_0=1, E_1 = 4 = R_1$, $E_2=14$ and $R_2 = \frac72$ we get: $R_s\leq \frac72$ (for $s\geq 2$) and
\be\label{eq:E_ineqa}
    E_{s+1}\leq\frac72 E_s,\qquad  (s\ge 1)
\ee
Iterating above from $E_1 = 4$ gives Eq.~\eqref{eq:ker_esa}.

Now for some rational \(\varrho>7/2\) let us define
\be\label{eq:I_rho}
        T_\varrho \equiv \sum_{s\ge1}E_s\varrho^{-s}.
\ee
From Eq.~\eqref{eq:ker_esa}
\begin{equation}
\begin{aligned}
T_\varrho
\le
\sum_{s\ge1}
4\left(\frac72\right)^{s-1}\varrho^{-s}
=
\frac{4}{\varrho}
\sum_{s\ge1}
\left(\frac{7}{2\varrho}\right)^{s-1}
=
\frac{4}{\varrho-\frac72},
\end{aligned}
\label{eq:T-rho-bound}
\end{equation}
which implies that $T_\varrho<1$ for $\varrho\geq 8$.

Now let us consider the following sum:
\be\label{eq:sum_c}
        c_n\equiv\sum_{s=0}^{n}E_sd_{n-s}.
\ee
Suppose that, for some \(\varrho>7/2\),
\be\label{eq:as-1a}
        |d_l|\ge \varrho |d_{l-1}|
        \qquad(1\le l\le n),
\ee
and suppose \(T_\varrho<1\). Then we can show that
\be\label{eq:c=d}
        {\rm{sgn}}(c_n)={\rm{sgn}}(d_n).
\ee
Moreover, the following inequalities also hold
\be\label{eq:ll0}
        (1-T_\varrho)|d_n|
        \le
        |c_n|
        \le
        (1+T_\varrho)|d_n|.
\ee
To see this let us note that Eq.~\eqref{eq:as-1a}, upon iteration, implies
\[
        |d_{n-s}|\le \varrho^{-s}|d_n|
        \qquad(1\le s\le n),
\]
which thereby leads to,
\be\label{eq:lemma1}
 \left|
 \sum_{s=1}^{n}E_s d_{n-s}
 \right|
 \leq
 \sum_{s=1}^{n}E_s|d_{n-s}|
 \leq
 |d_n|\sum_{s=1}^{n}E_s\varrho^{-s}
 \leq
 T_{\varrho}|d_n|.
\ee
since $E_s>0$. Now since \(T_\varrho<1\), the leading term in Eq.~\eqref{eq:sum_c}: \(E_0\,d_n=d_n\), dominates the remaining tail. Therefore \(c_n\) has the same sign as \(d_n\). The same estimate also gives
\[
        |c_n-d_n|\le T_\varrho |d_n|,
\]
which implies the two-sided bound given in Eq.~\eqref{eq:ll0}. Now from Eq.~\eqref{eq:ll0} we get:
\[
        |c_n|\ge (1-T_\varrho)|d_n|,
        \qquad
        |c_{n-1}|\le (1+T_\varrho)|d_{n-1}|.
\]
Therefore, we get:
\be\label{eq:rat_cn}
        \frac{|c_n|}{|c_{n-1}|}
        \ge
        \frac{1-T_\varrho}{1+T_\varrho}
        \frac{|d_n|}{|d_{n-1}|}
        \ge
        \varrho\frac{1-T_\varrho}{1+T_\varrho}.
\ee
If \(\varrho\ge10\), then \(T_\varrho\le T_{10}<2/3\), hence
\begin{equation}
 \frac{|c_n|}{|c_{n-1}|}
 \geq
 \varrho\frac{1-T_{\varrho}}{1+T_{\varrho}}
 >
 \frac{\varrho}{5}
 \geq2.
 \label{eq:rat_cn2}
\end{equation}
For later use, it is convenient to record a slightly more general,
local form of the preceding estimate. Fix an integer $k\geq1$ and
consider the single convolution coefficient
\be
 c_k=d_k+\sum_{s=1}^{k}E_s d_{k-s}.
\ee
To prove the sign and two-sided estimates for this particular
coefficient $c_k$, it is not necessary to assume that every adjacent
ratio $|d_l|/|d_{l-1}|$ is at least $\varrho$. It is sufficient to
assume directly that
\begin{equation}
 |d_{k-s}|
 \leq
 \varrho^{-s}|d_k|,
 \qquad
 1\leq s\leq k.
 \label{eq:local-backward-bound}
\end{equation}
Indeed, this gives
\be
 \left|
 \sum_{s=1}^{k}E_s d_{k-s}
 \right|
 \leq
 \sum_{s=1}^{k}E_s|d_{k-s}|
 \leq
 |d_k|\sum_{s=1}^{k}E_s\varrho^{-s}
 \leq
 T_\varrho|d_k|.
\ee
Consequently, if $T_\varrho<1$, then
\be
 \operatorname{sgn}(c_k)=\operatorname{sgn}(d_k),
 \qquad
 (1-T_\varrho)|d_k|
 \leq
 |c_k|
 \leq
 (1+T_\varrho)|d_k|.
\ee
If the local backward estimate
\eqref{eq:local-backward-bound} holds at both $k=n$ and $k=n-1$,
and if $|d_n|\geq\varrho|d_{n-1}|$, then
\be
 \frac{|c_n|}{|c_{n-1}|}
 \geq
 \varrho\frac{1-T_\varrho}{1+T_\varrho}.
\ee
This local formulation is useful when one individual adjacent ratio
is smaller than $\varrho$, but is compensated by a sufficiently large
neighbouring ratio.

Now let us consider the sum:
\be
        \tilde{c}_n = \sum_{s=0}^{n}E_s\,e_{n-s},
        \qquad
        e_l>0.
\ee
and let us suppose that $e_l\ge R\, e_{l-1}$ for some \(R>1\). Then, we can show that:
\be\label{eq:tilc}
        \tilde{c}_n\ge R \, \tilde{c}_{n-1}.
\ee
To show this, first let us note that
\[
        \tilde{c}_n
        \ge
        \sum_{l=1}^{n}E_{n-l}\,e_l,
\]
since $E_n>0$. Now using \(e_l\ge R\, e_{l-1}\), we get
\[
        \tilde{c}_n
        \ge
        R\sum_{l=1}^{n}E_{n-l}\,e_{l-1}=R\sum_{k=0}^{n-1}E_{n-1-k}\,e_k
        =
        R\,\tilde{c}_{n-1},
\]
where to get the first equality above we set $k=l-1$.

Next note that using Eq.~\eqref{eq:E_ineqa} and the fact that $E_0 = 1, \, E_1=4$, we can get the following weaker inequality:
\be
    E_{s+1}\leq 4 E_s, \qquad (s\geq 0)
\ee
Now if we assume,
\be\label{eq:xjE1}
        x_l<0,\qquad x_{l+1}>0,\qquad |x_{l+1}|>4|x_l|,
\ee
then for every \(n\ge l+1\),
\be\label{eq:xjE2a}
        E_{n-l}x_l+E_{n-l-1}x_{l+1}>0.
\ee
To see this, note that since \(E_{n-l}\le 4E_{n-l-1}\),
\[
\begin{aligned}
E_{n-l}x_l + E_{n-l-1}x_{l+1}
&\ge
-4E_{n-l-1}|x_l| + E_{n-l-1}|x_{l+1}|=
E_{n-l-1}\left(|x_{l+1}|-4|x_l|\right)>0.
\end{aligned}
\]
Next if we assume
\be\label{eq:xjE3}
        x_l,x_{l+1}<0,\qquad x_{l+2}>0,
\ee
and
\be\label{eq:xjE4}
        |x_{l+2}|>4|x_{l+1}|+16|x_l|,
\ee
then for every \(n\ge l+2\),
\be\label{eq:xjE5}
        E_{n-l}x_l + E_{n-l-1}x_{l+1} + E_{n-l-2}x_{l+2}>0.
\ee
To show the above, let us set \(s=n-l-2\). Then the three terms are
\be
        E_{s+2}x_l + E_{s+1}x_{l+1} + E_sx_{l+2}.
\ee
Since \(E_{s+1}\le 4E_s\) and \(E_{s+2}\le 16E_s\), we get
\be
\begin{aligned}
E_{s+2}x_l + E_{s+1}x_{l+1} + E_sx_{l+2}
&\ge
-16E_s|x_l| - 4E_s|x_{l+1}| + E_s|x_{l+2}|=
E_s\left(|x_{l+2}|-4|x_{l+1}|-16|x_l|\right)>0.
\end{aligned}
\ee

\subsection{A strengthening of the $\ell=0$ $\mA_1$-series growth estimate}
\label{app:A1-growth-200}

We now prove the stronger estimate required for the
$M^{(2)}<0$ branch. Let $N\in\mathbb Z_{>0}$ and write the Frobenius
recursion for the $\ell=0$, $\mA_1$ identity character in the form
\be\label{eq:A1-Frobenius-recursion}
 a^{(0)}_{0,n}(N)
 =
 \sum_{i=1}^{n}
 f_N(n,i)\,
 a^{(0)}_{0,n-i}(N).
\ee
In the alternating window $1\leq n\leq2N$, one has
$f_N(n,i)<0$. Define
\be
 u_n\equiv (-1)^n a^{(0)}_{0,n}(N),
 \qquad
 0\leq n\leq2N.
\ee
Eq.~\eqref{eq:A1-Frobenius-recursion} then becomes
\be\label{eq:A1-u-recursion}
 u_n
 =
 |f_N(n,1)|u_{n-1}
 +
 \sum_{i=2}^{n}
 (-1)^{i+1}|f_N(n,i)|u_{n-i}.
\ee
For the $\mA_1$ identity character,
\be\label{eq:A1-fN-one}
 |f_N(n,1)|
 =
 \frac{240N^2+64N+\frac{25}{4}-4n}
 {n\left(2N+\frac14-n\right)}.
\ee
The denominator in Eq.~\eqref{eq:A1-fN-one} is positive for
$1\leq n\leq2N$. Moreover,
\begin{align}
&240N^2+64N+\frac{25}{4}-4n
 -240n\left(2N+\frac14-n\right)
\nonumber\\
&\qquad =
240(n-N)^2-64(n-N)+\frac{25}{4}
\nonumber\\
&\qquad =
240\left(n-N-\frac{2}{15}\right)^2
+\frac{119}{60}>0.
\label{eq:A1-fN-240-proof}
\end{align}
It follows that
\be\label{eq:A1-fN-240}
 |f_N(n,1)|>240,
 \qquad
 1\leq n\leq2N.
\ee
The divisor-function estimates used in Eq.~(5.19) of \cite{Das:2025gto} also give
\be\label{eq:A1-fi-f1-ratio}
 \frac{|f_N(n,i)|}{|f_N(n,1)|}
 <
 4i^3,
 \qquad
 2\leq i\leq n\leq2N.
\ee

The case $n=1$ follows immediately from
Eq.~\eqref{eq:A1-fN-240}:
\be
 u_1=|f_N(1,1)|u_0>240u_0>200u_0.
\ee
Suppose inductively we can show
\be
 u_k>200u_{k-1}>0,
 \qquad
 1\leq k<n.
\ee
Then, the above implies
\be
 u_{n-i}
 <
 200^{-(i-1)}u_{n-1},
 \qquad
 2\leq i\leq n.
\ee
Using Eqs.~\eqref{eq:A1-u-recursion},
\eqref{eq:A1-fN-240} and \eqref{eq:A1-fi-f1-ratio}, we find
\begin{align}
u_n
&\geq
 |f_N(n,1)|u_{n-1}
 -
 \sum_{i=2}^{n}|f_N(n,i)|u_{n-i}
\nonumber\\
&>
240
\left(
1-4\sum_{i=2}^{\infty}
\frac{i^3}{200^{i-1}}
\right)u_{n-1}.
\label{eq:A1-growth-induction}
\end{align}
The standard generating-function identity
\be
 \sum_{i=2}^{\infty}i^3t^{i-1}
 =
 \frac{1+4t+t^2}{(1-t)^4}-1,
 \qquad |t|<1,
\ee
evaluated at the positive real value $t=1/200$, gives
\[
 \sum_{i=2}^{\infty}\frac{i^3}{200^{i-1}}
 =
 \frac{
 1+\frac{4}{200}+\frac{1}{200^2}
 }{
 \left(1-\frac{1}{200}\right)^4
 }
 -1.
\]
Consequently,
\be
240\left(
1-4\sum_{i=2}^{\infty}
\frac{i^3}{200^{i-1}}
\right)
=
\frac{315128641200}{1568239201}
>200.
\ee
Therefore the induction closes and we have shown
\be\label{eq:A1-growth-200}
 \left|a^{(0)}_{0,n}(N)\right|
 >
 200
 \left|a^{(0)}_{0,n-1}(N)\right|,
 \qquad
 1\leq n\leq2N.
\ee

\subsection{Proof of signs and growth of $q$-series coefficients}\label{app:sgn_an_dn}

\paragraph{Identity coefficients $a_{0,n}^{(2)}$:}
Let us consider the sum in Eq.~\eqref{eq:Mpos-id-conv}. If we can show:
\be\label{eq:as-1}
        |d_l|\ge \varrho |d_{l-1}|
        \qquad(1\le l\le n),
\ee
for some \(\varrho\geq 8\), then using Eq.~\eqref{eq:c=d} we get: $\text{sgn}(a_{0,l}^{(2)}(M^{(2)}))=\text{sgn}(d_l)$. To show the above, let us consider the ratio
\[
\frac{|d_l|}{|d_{l-1}|}
=
\frac{\left|l-M^{(2)}-\frac{19}{24}\right|}
{\left|l-1-M^{(2)}-\frac{19}{24}\right|}
\frac{|a^{(0)}_{1,l}(-M^{(2)}-1)|}{|a^{(0)}_{1,l-1}(-M^{(2)}-1)|},
\]
for \(1\le l\le2M^{(2)}+1\). The smallest possible value of the above ratio occurs at the crossover point \(l=M^{(2)}+1\). So, using Eq.~\eqref{eq:N<0_nid_2} we get:
\[
        \frac{|d_{M^{(2)}+1}|}{|d_{M^{(2)}}|}
        >
        \frac{5/24}{19/24}\cdot45
        =
        \frac{225}{19}
        >10.
\]
Thus
\begin{equation}
        |d_l|>\frac{225}{19}|d_{l-1}|
        \qquad(1\le l\le2M^{(2)}+1).
\label{eq:Mpos-id-raw-growth}
\end{equation}
Now setting $\varrho = \frac{225}{19}>8$, we get Eq.~\eqref{eq:as-1}. Hence, we finally get,
\[
        \operatorname{sgn}\left(a^{(2)}_{0,n}(M^{(2)})\right)
        =
        \operatorname{sgn}(d_n)
        \qquad(1\le n\le2M^{(2)}+1).
\]
Hence
\be
\begin{split}
        &\operatorname{sgn}\left(a^{(2)}_{0,n}(M^{(2)})\right)
        =
        (-1)^n
        \qquad(1\le n\le M^{(2)}), \\
        &\operatorname{sgn}\left(a^{(2)}_{0,n}(M^{(2)})\right)
        =
        (-1)^{n+1}
        \quad(M^{(2)}+1\le n\le2M^{(2)}+1)
\end{split}        
\ee
In particular, $a^{(2)}_{0,2M^{(2)}+1}(M^{(2)})>0$. Next let us prove that all later coefficients are positive. From Eq.~\eqref{eq:a1ns}, we see that \(a^{(0)}_{1,l}(-M^{(2)}-1)<0\)  for \(l\ge2M^{(2)}+1\) while $l-M^{(2)}-\frac{19}{24}>0$. Therefore, by Eq.~\eqref{eq:Mpos-id-raw},
\[
        d_l>0
        \qquad(l\ge2M^{(2)}+1).
\]
Thus the only negative terms $d_l$ occur in the finite region \(0\le l\le2M^{(2)}\). 

Now let us pair each negative term with the next positive term. For pairs
\be\label{eq:pair}
        d_l<0,\qquad d_{l+1}>0,
\ee
from the ratio estimate Eq.~\eqref{eq:Mpos-id-raw-growth} we can get a weaker inequality: $|d_{l+1}|>4|d_l|$ (since $\varrho>4$). Then we can show that:
\be\label{eq:xjE2}
    E_{n-l}d_l + E_{n-l-1}d_{l+1}>0
\ee
for every $n\geq l+1$ (see appendix \ref{app:estimates_log} for details). Thus, by Eq.~\eqref{eq:xjE2}, we have, each such pair contributing positively to every coefficient \(a^{(2)}_{0,n}(M^{(2)})\) once both terms of the pair as in Eq.~\eqref{eq:pair} are present in Eq.~\eqref{eq:Mpos-id-conv}.

There is only one possible obstruction to the above: if \(M^{(2)}\) is odd, then the crossover block has two consecutive negative terms:
\[
        d_{M^{(2)}}<0,\qquad d_{M^{(2)}+1}<0,\qquad d_{M^{(2)}+2}>0.
\]
However,
\[
        \frac{|d_{M^{(2)}+1}|}{|d_{M^{(2)}}|}
        >
        \frac{225}{19}, \quad \text{and} \quad \frac{|d_{M^{(2)}+2}|}{|d_{M^{(2)}+1}|}
        >
        \frac{29/24}{5/24}\cdot45
        =
        261.
\]
Therefore,
\[
        |d_{M^{(2)}+2}|
        >
        4|d_{M^{(2)}+1}|+16|d_{M^{(2)}}|.
\]
With the above we can show the following three-term part result:
\be\label{eq:Mpos-three-term-block}
        E_{n-l}d_l + E_{n-l-1}d_{l+1} + E_{n-l-2}d_{l+2}>0,
\ee
for every $n\geq l+2$ with $l=M^{(2)}$. Hence, by above result, the whole crossover block contributes positively. All unpaired terms in the tail are positive. Therefore
\[
        a^{(2)}_{0,n}(M^{(2)})>0
        \qquad(n\ge2M^{(2)}+1).
\]

Now to obtain a geometric growth for the coefficients in the range: $1\le n\le2M^{(2)}+1$, we set $\varrho = \frac{225}{19}>10$. Then, we get from Eqs.~\eqref{eq:rat_cn} and \eqref{eq:rat_cn2}
\begin{equation}
        \left|a^{(2)}_{0,n}(M^{(2)})\right|
        >
        2\left|a^{(2)}_{0,n-1}(M^{(2)})\right|
        \qquad(1\le n\le2M^{(2)}+1).
\label{eq:Mpos-id-supergrowtha}
\end{equation}

\paragraph{Geometric growth for non-identity coefficients $a_{1,n}^{(2)}$:}
To obtain a geometric growth for the non-identity coefficients, consider the pre-factor in Eq.~\eqref{eq:defn_el} which is positive and increasing in \(l\). Now we have from Eq.~\eqref{eq:N<0_id_1}
\[
        a^{(0)}_{0,l}(-M^{(2)}-1)
        \ge
        4\,a^{(0)}_{0,l-1}(-M^{(2)}-1)
        \qquad(1\le l\le2M^{(2)}+2).
\]
Therefore
\be\label{eq:assump_el}
        e_l\ge 4 \, e_{l-1}
        \qquad(1\le l\le2M^{(2)}+2),
\ee
Hence, by using Eq.~\eqref{eq:tilc} we get
\be\label{eq:Mpos-nonid-supergrowtha}
        a^{(2)}_{1,n}(M^{(2)})
        \ge
        4\,a^{(2)}_{1,n-1}(M^{(2)})
        \qquad(1\le n\le2M^{(2)}+2).
\ee

\subsection{The case of $M^{(2)}<0$}
\label{app:Mneg-proof}

Now let us consider the case $M^{(2)}<0$. The corresponding $\ell=0$ source parameter is $N=|M^{(2)}|-1$.

\paragraph{Identity component:}
The first relation in Eq.~\eqref{eq:Theta-M-negative} gives
\be\label{eq:Mneg-id-convolution}
 a^{(2)}_{0,n}(M^{(2)})
 =
 \sum_{s=0}^{n}
 E_s
 \left(
 n-s+|M^{(2)}|-\frac{19}{24}
 \right)
 a^{(0)}_{1,n-s}(|M^{(2)}|-1).
\ee
Every factor in every summand is positive. For $|M^{(2)}|\geq2$, this follows
from Eq.~\eqref{eq:N>0_nid_0}. For $|M^{(2)}|=1$, the source is the
non-identity character of the admissible $\mA_{1,1}$ theory, whose
coefficients are likewise positive. Since
\[
 n-s+|M^{(2)}|-\frac{19}{24}>0,
 \qquad
 (0\leq s\leq n),
\]
Eq.~\eqref{eq:Mneg-id-convolution} proves
Eq.~\eqref{eq:Mneg-id-positive}.

\paragraph{Non-identity component for $|M^{(2)}|\geq2$:}
The second relation in Eq.~\eqref{eq:Theta-M-negative} gives
\be\label{eq:Mneg-nonid-convolution}
 a^{(2)}_{1,n}(M^{(2)})
 =
 \sum_{s=0}^{n}E_s d_{n-s},
 \qquad
 d_j\equiv
 \left(
 |M^{(2)}|-\frac{23}{24}-j
 \right)
 a^{(0)}_{0,j}(|M^{(2)}|-1).
\ee
In the alternating source window, let us write
\[
 a^{(0)}_{0,j}(|M^{(2)}|-1)=(-1)^j u_j,
 \qquad
 u_j>0,
 \qquad
 (0\leq j\leq2|M^{(2)}|-2).
\]
By Eq.~\eqref{eq:A1-growth-200},
\be\label{eq:Mneg-source-200}
 u_j>200u_{j-1},
 \qquad
 (1\leq j\leq2|M^{(2)}|-2).
\ee
Moreover,
\[
 a^{(0)}_{0,j}(|M^{(2)}|-1)>0,
 \qquad
 (j\geq2|M^{(2)}|-1).
\]
It follows from Eq.~\eqref{eq:Mneg-nonid-convolution} that
\be
 \operatorname{sgn}(d_j)=(-1)^j,
 \qquad
 (0\leq j\leq |M^{(2)}|-1),
\ee
\be
 \operatorname{sgn}(d_j)=(-1)^{j+1},
 \qquad
 (|M^{(2)}|\leq j\leq2|M^{(2)}|-2),
\ee
and
\be
 d_j<0,
 \qquad
 (j\geq2|M^{(2)}|-1).
\ee

For $1\leq j\leq2|M^{(2)}|-2$,
\[
 \frac{|d_j|}{|d_{j-1}|}
 >
 200
 \left|
 \frac{|M^{(2)}|-j-\frac{23}{24}}
 {|M^{(2)}|-j+\frac{1}{24}}
 \right|.
\]
The only weak step occurs at $j=|M^{(2)}|-1$. More explicitly,
\begin{align}
 \frac{|d_j|}{|d_{j-1}|}
 &>
 \frac{5000}{49}>100,
 &&(1\leq j\leq |M^{(2)}|-2),
 \label{eq:Mneg-pre-edge}
 \\
 \frac{|d_{|M^{(2)}|-1}|}{|d_{|M^{(2)}|-2}|}
 &>8,
 \label{eq:Mneg-weak-edge}
 \\
 \frac{|d_{|M^{(2)}|}|}{|d_{|M^{(2)}|-1}|}
 &>4600,
 \label{eq:Mneg-large-edge}
 \\
 \frac{|d_j|}{|d_{j-1}|}
 &>200,
 &&(|M^{(2)}|+1\leq j\leq2|M^{(2)}|-2).
 \label{eq:Mneg-post-edge}
\end{align}
In the first line, $5000/49$ is the weakest value in the indicated
range; it is attained at the step $j=|M^{(2)}|-2$ (when \(\lvert M^{(2)}\rvert\ge3\)) at the level of the stated
lower bound.

All the adjacent ratios are therefore larger than $8$. From
Eq.~\eqref{eq:T-rho-bound},
\[
 T_8\leq\frac89<1.
\]
Consequently, Eqs.~\eqref{eq:c=d} and \eqref{eq:ll0} imply
\be
 \operatorname{sgn}\left(a^{(2)}_{1,n}(M^{(2)})\right)
 =
 \operatorname{sgn}(d_n),
 \qquad
 (0\leq n\leq2|M^{(2)}|-2).
\ee
This proves the two finite sign windows in
Eq.~\eqref{eq:Mneg-nonid-sign-final}.

We next prove the finite-window growth estimate. From
Eq.~\eqref{eq:T-rho-bound},
\[
 T_{100}
 \leq
 \frac{8}{193}
 <
 \frac1{20}.
\]
For $1\leq n\leq |M^{(2)}|-2$, all backward ratios from $d_n$ are larger
than $100$. Equations~\eqref{eq:rat_cn} and
\eqref{eq:rat_cn2} therefore give
\be
 \frac{|a^{(2)}_{1,n}(M^{(2)})|}
 {|a^{(2)}_{1,n-1}(M^{(2)})|}>2.
\ee

The step $n=|M^{(2)}|-1$ requires a separate estimate. From
Eqs.~\eqref{eq:Mneg-pre-edge} and
\eqref{eq:Mneg-weak-edge},
\be
 |d_{|M^{(2)}|-1-s}|
 \leq
 \frac{|d_{|M^{(2)}|-1}|}{8\,(100^{s-1})},
 \qquad
 (1\leq s\leq |M^{(2)}|-1).
\ee
Hence
\begin{align}
\left|
 a^{(2)}_{1,|M^{(2)}|-1}(M^{(2)})-d_{|M^{(2)}|-1}
\right|
&\leq
\frac{|d_{|M^{(2)}|-1}|}{8}
\sum_{s=1}^{|M^{(2)}|-1}
\frac{E_s}{100^{s-1}}
\nonumber\\
&<
\frac{100}{8}T_{100}|d_{|M^{(2)}|-1}|
<
\frac58|d_{|M^{(2)}|-1}|.
\label{eq:Mneg-mminusone-tail}
\end{align}
Thus
\be
 |a^{(2)}_{1,|M^{(2)}|-1}(M^{(2)})|
 >
 \frac38|d_{|M^{(2)}|-1}|.
\ee
On the other hand,
\be
 |a^{(2)}_{1,|M^{(2)}|-2}(M^{(2)})|
 <
 (1+T_{100})|d_{|M^{(2)}|-2}|
 <
 \frac{21}{20}|d_{|M^{(2)}|-2}|.
\ee
Using Eq.~\eqref{eq:Mneg-weak-edge}, we obtain
\be\label{eq:Mneg-mminusone-growth}
 \frac{|a^{(2)}_{1,|M^{(2)}|-1}(M^{(2)})|}
 {|a^{(2)}_{1,|M^{(2)}|-2}(M^{(2)})|}
 >
 \frac{3/8}{21/20}8
 =
 \frac{20}{7}>2.
\ee

The step $n=|M^{(2)}|$ must also be treated separately. The two exceptional
raw ratios satisfy
\[
 8\cdot4600>100^2.
\]
Consequently,
\be
 |d_{|M^{(2)}|-s}|
 \leq
 100^{-s}|d_{|M^{(2)}|}|,
 \qquad
 (1\leq s\leq |M^{(2)}|),
\ee
and therefore
\be
 |a^{(2)}_{1,|M^{(2)}|}(M^{(2)})|
 >
 (1-T_{100})|d_{|M^{(2)}|}|
 >
 \frac{19}{20}|d_{|M^{(2)}|}|.
\ee
At the preceding edge,
Eq.~\eqref{eq:Mneg-mminusone-tail} also gives
\be
 |a^{(2)}_{1,|M^{(2)}|-1}(M^{(2)})|
 <
 \left(
 1+\frac{100}{8}T_{100}
 \right)|d_{|M^{(2)}|-1}|
 <
 \frac{13}{8}|d_{|M^{(2)}|-1}|.
\ee
Using Eq.~\eqref{eq:Mneg-large-edge}, we find
\be\label{eq:Mneg-m-growth}
 \frac{|a^{(2)}_{1,|M^{(2)}|}(M^{(2)})|}
 {|a^{(2)}_{1,|M^{(2)}|-1}(M^{(2)})|}
 >
 \frac{(19/20)\,4600}{13/8}
 =
 \frac{34960}{13}>2.
\ee

For $|M^{(2)}|+1\leq n\leq2|M^{(2)}|-2$, every cumulative backward product satisfies
\be
 |d_{k-s}|
 \leq
 100^{-s}|d_k|,
 \qquad
 (1\leq s\leq k, \, \, \text{ with }
 \quad
 k=n,n-1).
\ee
Indeed, a product crossing the weak edge
\eqref{eq:Mneg-weak-edge} also contains the adjacent large factor
\eqref{eq:Mneg-large-edge}, and
$8\cdot4600>100^2$. The local form
\eqref{eq:local-backward-bound} therefore applies at both endpoints.
Together with Eqs.~\eqref{eq:Mneg-mminusone-growth} and
\eqref{eq:Mneg-m-growth}, this proves
Eq.~\eqref{eq:Mneg-nonid-growth}.

It remains to establish the eventual negative sign. For
$j\geq2|M^{(2)}|-1$, one has $d_j<0$. Pair each positive raw term in the
finite interval $0\leq j\leq2|M^{(2)}|-2$ with the following negative raw
term. Every ordinary pair satisfies
\be
 d_j>0,
 \qquad
 d_{j+1}<0,
 \qquad
 |d_{j+1}|>4|d_j|.
\ee
The two-term pairing estimate, with all signs reversed, shows that
each such pair contributes negatively once both terms are present.

If $|M^{(2)}|$ is odd, the crossover instead contains the three-term block
\be
 d_{|M^{(2)}|-1}>0,
 \qquad
 d_{|M^{(2)}|}>0,
 \qquad
 d_{|M^{(2)}|+1}<0.
\ee
Here
\be
 \frac{|d_{|M^{(2)}|}|}{|d_{|M^{(2)}|-1}|}>4600,
 \qquad
 \frac{|d_{|M^{(2)}|+1}|}{|d_{|M^{(2)}|}|}
 >
 \frac{47}{23}\,200
 =
 \frac{9400}{23}.
\ee
These estimates imply
\be
 |d_{|M^{(2)}|+1}|>4|d_{|M^{(2)}|}|+16|d_{|M^{(2)}|-1}|.
\ee
The three-term pairing estimate, again with all signs reversed, shows
that this block also contributes negatively. All remaining unpaired
raw terms are negative. Therefore
\be
 a^{(2)}_{1,n}(M^{(2)})<0,
 \qquad
 n\geq2|M^{(2)}|-1.
\ee
This completes the proof of
Eq.~\eqref{eq:Mneg-nonid-sign-final} for $|M^{(2)}|\geq2$.

\paragraph{The special case $|M^{(2)}|=1$:}
The source is the admissible $\mA_{1,1}$ identity character,
\be
 \chi^{(0)}_{0,0}(q)
 =
 q^{-1/24}
 \left(
 1+3q+4q^2+\cdots
 \right).
\ee
The raw sequence is
\be
 d_j=
 \left(
 \frac1{24}-j
 \right)
 a^{(0)}_{0,j}(0).
\ee
Thus $d_0>0$, while $d_j<0$ for every $j\geq1$, and
\be
 \frac{|d_1|}{d_0}
 =
 \frac{(23/24)\,3}{1/24}
 =
 69>4.
\ee
The two-term pairing of $d_0$ with $d_1$, together with the remaining
negative raw tail, gives
\be
 a^{(2)}_{1,n}(-1)<0,
 \qquad
 n\geq1.
\ee
This agrees with Eq.~\eqref{eq:Mneg-nonid-sign-final}; the range in
Eq.~\eqref{eq:Mneg-nonid-growth} is empty when $|M^{(2)}|=1$.

\bibliography{two_char}
\bibliographystyle{JHEP}

\end{document}